\RequirePackage{amsmath}
\RequirePackage{fix-cm}
\documentclass[smallextended]{svjour3hack}       % onecolumn (second format)
\usepackage[]{graphicx}
\usepackage[table]{xcolor}

\makeatletter
\def\maxwidth{ %
  \ifdim\Gin@nat@width>\linewidth
    \linewidth
  \else
    \Gin@nat@width
  \fi
}
\makeatother

\definecolor{fgcolor}{rgb}{0.345, 0.345, 0.345}
\newcommand{\hlnum}[1]{\textcolor[rgb]{0.686,0.059,0.569}{#1}}%
\newcommand{\hlstr}[1]{\textcolor[rgb]{0.192,0.494,0.8}{#1}}%
\newcommand{\hlcom}[1]{\textcolor[rgb]{0.678,0.584,0.686}{\textit{#1}}}%
\newcommand{\hlopt}[1]{\textcolor[rgb]{0,0,0}{#1}}%
\newcommand{\hlstd}[1]{\textcolor[rgb]{0.345,0.345,0.345}{#1}}%
\newcommand{\hlkwa}[1]{\textcolor[rgb]{0.161,0.373,0.58}{\textbf{#1}}}%
\newcommand{\hlkwb}[1]{\textcolor[rgb]{0.69,0.353,0.396}{#1}}%
\newcommand{\hlkwc}[1]{\textcolor[rgb]{0.333,0.667,0.333}{#1}}%
\newcommand{\hlkwd}[1]{\textcolor[rgb]{0.737,0.353,0.396}{\textbf{#1}}}%

\usepackage{framed}
\makeatletter
\newenvironment{kframe}{%
 \def\at@end@of@kframe{}%
 \ifinner\ifhmode%
  \def\at@end@of@kframe{\end{minipage}}%
  \begin{minipage}{\columnwidth}%
 \fi\fi%
 \def\FrameCommand##1{\hskip\@totalleftmargin \hskip-\fboxsep
 \colorbox{shadecolor}{##1}\hskip-\fboxsep
     \hskip-\linewidth \hskip-\@totalleftmargin \hskip\columnwidth}%
 \MakeFramed {\advance\hsize-\width
   \@totalleftmargin\z@ \linewidth\hsize
   \@setminipage}}%
 {\par\unskip\endMakeFramed%
 \at@end@of@kframe}
\makeatother

\definecolor{shadecolor}{rgb}{.97, .97, .97}
\definecolor{messagecolor}{rgb}{0, 0, 0}
\definecolor{warningcolor}{rgb}{1, 0, 1}
\definecolor{errorcolor}{rgb}{1, 0, 0}
\newenvironment{knitrout}{}{} % an empty environment to be redefined in TeX

\usepackage{alltt}
\usepackage[utf8]{inputenc}
\usepackage[T1]{fontenc}
\usepackage{amssymb,array}
\usepackage{booktabs}
\usepackage{caption}
\usepackage{cite}
\usepackage{colortbl}
\usepackage{comment}
\usepackage{float}
\usepackage{graphicx}

\usepackage{multirow}
\usepackage{pdflscape}
\usepackage{tabu}
\usepackage{threeparttable}
\usepackage[normalem]{ulem}
\usepackage{wrapfig}
\usepackage{pdfpages}
\usepackage{placeins}

\usepackage{scalerel,stackengine}
\usepackage{subcaption}
\stackMath
\newcommand\reallywidehat[1]{%
\savestack{\tmpbox}{\stretchto{%
  \scaleto{%
    \scalerel*[\widthof{\ensuremath{#1}}]{\kern-.6pt\bigwedge\kern-.6pt}%
    {\rule[-\textheight/2]{1ex}{\textheight}}%WIDTH-LIMITED BIG WEDGE
  }{\textheight}%
}{0.5ex}}%
\stackon[1pt]{#1}{\tmpbox}%
}

\PassOptionsToPackage{hyphens}{url}
\usepackage[hypertexnames=false,colorlinks=true,breaklinks=true,bookmarks=true,urlcolor=blue,citecolor=blue,linkcolor=blue,bookmarksopen=false,draft=false]{hyperref}

\title{An R Package for generating covariance matrices \\ for maximum-entropy sampling from precipitation chemistry data}

\titlerunning{MESgenCov}        % if too long for running head

\author{Hessa Al-Thani \and Jon Lee}
\institute{H. Al-Thani\at
              Department of Industrial and Operations Engineering, Univ. of Michigan, Ann Arbor, MI 48105 USA  \\
              \email{hessakh@umich.edu}           %  \\
           \and
           J. Lee \at
          Department of Industrial and Operations Engineering, Univ. of Michigan, Ann Arbor, MI 48105 USA  \\
              \email{jonxlee@umich.edu}           %
}

\date{\today}

\IfFileExists{upquote.sty}{\usepackage{upquote}}{}

\journalname{SN Operations Research Forum}

\begin{document}

\maketitle

\begin{abstract}
  We present an open-source R package (\textbf{MESgenCov} v 0.1.0) for temporally fitting multivariate precipitation chemistry data and extracting a covariance matrix
for use in the MESP (maximum-entropy sampling problem).
 We provide multiple functionalities for modeling and model assessment. The package is
  tightly coupled with  NADP/NTN (National Atmospheric Deposition Program / National Trends Network) data from  their set of 379
 monitoring sites, 1978--present. The user specifies the sites, chemicals, and time period desired, fits an appropriate user-specified univariate model for each site and chemical selected, and the package produces a covariance matrix for use by MESP algorithms.
\end{abstract}

\section*{Introduction}

The MESP (maximum-entropy sampling problem) (see \cite{ShewryWynn,SebWynn,FedorovLee,LeeEnv}) has been applied to many domains where the objective is to determine a "most informative" subset $Y_S$, of pre-specified size $s=|S|>0$, from a Gaussian random vecor $Y_N$, $|N|=n>s$. Information is
typically measured by (differential) entropy.
Generally, we assume that $Y_N$ has a joint Gaussian distribution with mean vector $\mu$ and covariance matrix $C$. Up to constants, the  entropy of $Y_S$ is the log of the determinant of the principle submatrix $C[S,S]$. So, the MESP seeks to maximize the (log) determinant of $C[S,S]$, for some $S \subseteq N$ with $|S|=s$.

The MESP is NP-hard (see \cite{KLQ}), and
there has been considerable work on algorithms aimed at exact solutions for problems of moderate size;
see \cite{KLQ,LeeConstrained,AFLW_Using,LeeWilliams_ILP,HLW,AnstreicherLee_Masked,BurerLee,AnstreicherBQPEntropy,linx,Mixing}.
All of this algorithmic work is based on a branch-and-bound framework introduced in \cite{KLQ}, and the bulk of the contributions in  these references is on different methods for upper bounding the optimal value.
This work has been developed and validated in the context of a very small number of data sets,
despite the fact that of course multivariate data is widely available. The reason for this shortcoming
is that despite all
of the raw  multivariate data that is available, it is not a simple matter to turn this data into meaningful
covariance matrices for Gaussian random variables.

Our goal with the R package (\textbf{MESgenCov} v 0.1.0) that we have developed is to provide such  a link --- between readily available raw environmental-monitoring
data and covariance matrices suitable for the MESP --- in the context of environmental monitoring. Our work fits squarely into
recent efforts to better exploit massive amounts of available data for
mathematical-programming approaches to decision problems. Even if we have
reliable raw data, we can only make good decisions if we have a
means to prepare that data so that we
can populate our mathematical-programming models in such a was as to
meet required assumptions.

We note that another R package of interest is
\cite{EnviroStat}: "EnviroStat provides functions for spatio-temporal modeling of environmental processes and designing monitoring networks for them based on an approach described in \cite{LeZidek_book}".

In \S\ref{sec:EnvMon}, we discuss application of the MESP to environmental monitoring
and the NADP/NTN (National Atmospheric Deposition Program / National Trends Network) data environment. In \S\ref{sec:method}, we describe our methodology.
In \S\ref{sec:Imp}, we describe our R package (\textbf{MESgenCov} v 0.1.0).
In \S\ref{sec:Conc}, we make some concluding remarks.

\section{Environmental monitoring and NADP/NTN data}\label{sec:EnvMon}
A key area of application for the MESP has been in environmental monitoring (see \cite{Zidek1,Zidek2,Zidek3}, for example).
The idea is that precipitation is collected at many sites, and its chemistry is analyzed. This is costly, and it is
a natural question as to whether a subset of the sites might yield data without much loss of information (as measured by entropy). But it is a challenge to process the raw data in
such a way that multivariate normality is achieved, because only then are the model of the MESP and
its related algorithms  applicable.

The NADP maintains the NTN (see \cite{NADPNTN}); this network measures the chemistry (i.e., ammonium,
calcium,
chloride,
hydrogen,
 magnesium,
 nitrate,
pH,
 potassium,
  sodium, and
sulfate) of precipitation at 379
%\color{blue}  if the earlier one was changed, change this too \color{black}
monitoring sites across the US, with some data available as far back as 1978;
at present, 255 sites are active.

Our R package is tightly coupled with this precipitation and chemistry data.
We are interested in instances of the MESP where $n$ \emph{user-specified} site/chemical pairs comprise $N$.
Precipitation data (measured in L)
are available
on a daily basis, and chemical concentrations (measured in mg/L) are available on a weekly basis.
%%add a bit on generalized getCov after finishing the function
%\section{NADP data description}\label{sec:NADPdata}
 These datasets are available in the packages \texttt{EnvMonDataConc} and \texttt{EnvMonDataPre}.
 They can be installed and loaded using the \texttt{devtools} library.
 If not already installed,  the  \texttt{devtools} library can be installed and loaded as follows:
\begin{knitrout}
\definecolor{shadecolor}{rgb}{0.969, 0.969, 0.969}\color{fgcolor}\begin{kframe}
\begin{alltt}
 \hlcom{#Install and load devtools}
 \hlkwb{>}   \hlkwd{install.packages}\hlstd{(}\hlstr{"devtools"}\hlstd{)}
 \hlkwb{>}   \hlkwd{library}\hlstd{(}\hlstr{devtools}\hlstd{)}
\end{alltt}
\end{kframe}
\end{knitrout}
%Then we can use the \texttt{devtools} function \verb;install_github(); to download the \textbf{MESgenCov}  package from GitHub.
%
\noindent Then we can  install the data packages
(from GitHub) and load them:
 \begin{knitrout}
\definecolor{shadecolor}{rgb}{0.969, 0.969, 0.969}\color{fgcolor}\begin{kframe}
\begin{alltt}
\hlcom{#Install data packages}
 \hlkwb{>}   \hlkwd{install_github}\hlstd{(}\hlstr{"hessakh/EnvMonDataConc"}\hlstd{)}
 \hlkwb{>}   \hlkwd{install_github}\hlstd{(}\hlstr{"hessakh/EnvMonDataPre"}\hlstd{)}
\end{alltt}
%\end{kframe}
%\end{knitrout}
%\begin{knitrout}
%\definecolor{shadecolor}{rgb}{0.969, 0.969, 0.969}\color{fgcolor}\begin{kframe}
\begin{alltt}
\hlcom{#Load data packages}
 \hlkwb{>}   \hlkwd{library}\hlstd{(EnvMonDataConc)}
 \hlkwb{>}   \hlkwd{data}\hlstd{(}\hlstr{"weeklyConc"}\hlstd{)}
 \hlkwb{>}   \hlkwd{library}\hlstd{(EnvMonDataPre)}
 \hlkwb{>}   \hlkwd{data}\hlstd{(}\hlstr{"preDaily"}\hlstd{)}
\end{alltt}
\end{kframe}
\end{knitrout}

A full description of the daily and weekly precipitation data
appears in \hyperref[appA]{Appendix A},
derived from
\url{http://nadp.slh.wisc.edu/data/ntn/meta/ntn-daily-Meta.pdf} and
\url{http://nadp.slh.wisc.edu/data/ntn/meta/ntn-weekly-Meta.pdf},
courtesy of the
 NADP\footnote{National Atmospheric Deposition Program (NRSP-3). 2019.
NADP Program Office, Wisconsin State Laboratory of Hygiene,
465 Henry Mall, Madison, WI 53706.}

Small snapshots of the data can easily be viewed.
For example, we can output the first 6 rows and first 5 columns of the weekly raw data.

\begin{knitrout}
\definecolor{shadecolor}{rgb}{0.969, 0.969, 0.969}\color{fgcolor}\begin{kframe}
\begin{alltt}
\hlcom{#Display part of the weeklyConc data frame}
 \hlstd{weeklyConc[}\hlnum{1}\hlopt{:}\hlnum{6}\hlstd{,}\hlnum{1}\hlopt{:}\hlnum{5}\hlstd{]}
\end{alltt}
\begin{verbatim}
   siteID              dateon             dateoff yrmonth    ph
 1   AB32 2016-09-13 18:40:00 2016-09-20 15:10:00  201609 -9.00
 2   AB32 2016-09-20 15:15:00 2016-09-28 16:00:00  201609 -9.00
 3   AB32 2016-09-28 16:00:00 2016-10-05 16:55:00  201610  6.56
 4   AB32 2016-10-05 16:55:00 2016-10-11 17:00:00  201610 -9.00
 5   AB32 2016-10-11 17:00:00 2016-10-18 20:00:00  201610 -9.00
 6   AB32 2016-10-18 20:00:00 2016-10-25 18:00:00  201610  4.73
\end{verbatim}
\end{kframe}
\end{knitrout}

\noindent Note that missing concentration values are coded as -9.00.

\section{Our methodology}\label{sec:method}
  \subsection{NADP/NTN data processing}
  We process the raw NADP/NTN data in a similar way to earlier uses in the context of the MESP in the field of environmental statistics (see  \cite{Zidek3}).

 We calculate the level of a chemical's concentration by summing weekly quantities (mg) of the chemical, over a month,
 and dividing the monthly total by total precipitation (L), over dates in that month, to get monthly values of chemical concentration (mg/L).   We use monthly concentrations instead of the available weekly concentrations because there is a large proportion of missing data for individual weeks compared to full months. Furthermore the univariate models were better at predicting average monthly concentrations than they were are at predicting weekly concentrations.

 For a given monitoring site, chemical, and month $t=0,1,\ldots,T-1$, let
      \begin{align*}
        W(t) &:= \text{set of weeks in month $t$,}\\
        D(w) &:= \text{set of days in week $w$,} \\
        c_w  &:= \text{recorded chemical concentration (mg/L) for week $w$}\\
             &\phantom{:=} \text{ ($c_w=*$ denotes an unrecorded value),}\\
        p_d  &:= \text{recorded precipitation quantity (L) for day $d$,}\\
        p_w  &:= \text{precipitation quantity (L) for week $w$; $\textstyle p_w=\sum_{d\in D(w)}p_d$.}
    \end{align*}
Then the chemical concentration (mg/L) for month $t$ is calculated as
\[
y(t):=\frac{\sum_{w\in W(t) : c_w \not=*} p_w c_w}
{\sum_{w\in W(t) : c_w \not=*} \sum_{d\in w} p_d}.
\]
It should be noted that when there is no weekly value available for the chemical quantity,
    we do not use the precipitation values for any of the days in such a week
    (so as to not artificially dilute the chemical concentration level for the month).

Next, we fit a temporal model to $\log(y(t))$, which is a rather standard method
for handling heavy-tailed distributions. We note that our data has no zero concentrations,
so there is no issue of "$\log(0)$".

\begin{figure}[t!]
\centering
\begin{tabular}{cc}
      % after \\: \hline or \cline{col1-col2} \cline{col3-col4} ...
      \includegraphics[width=0.47\textwidth]{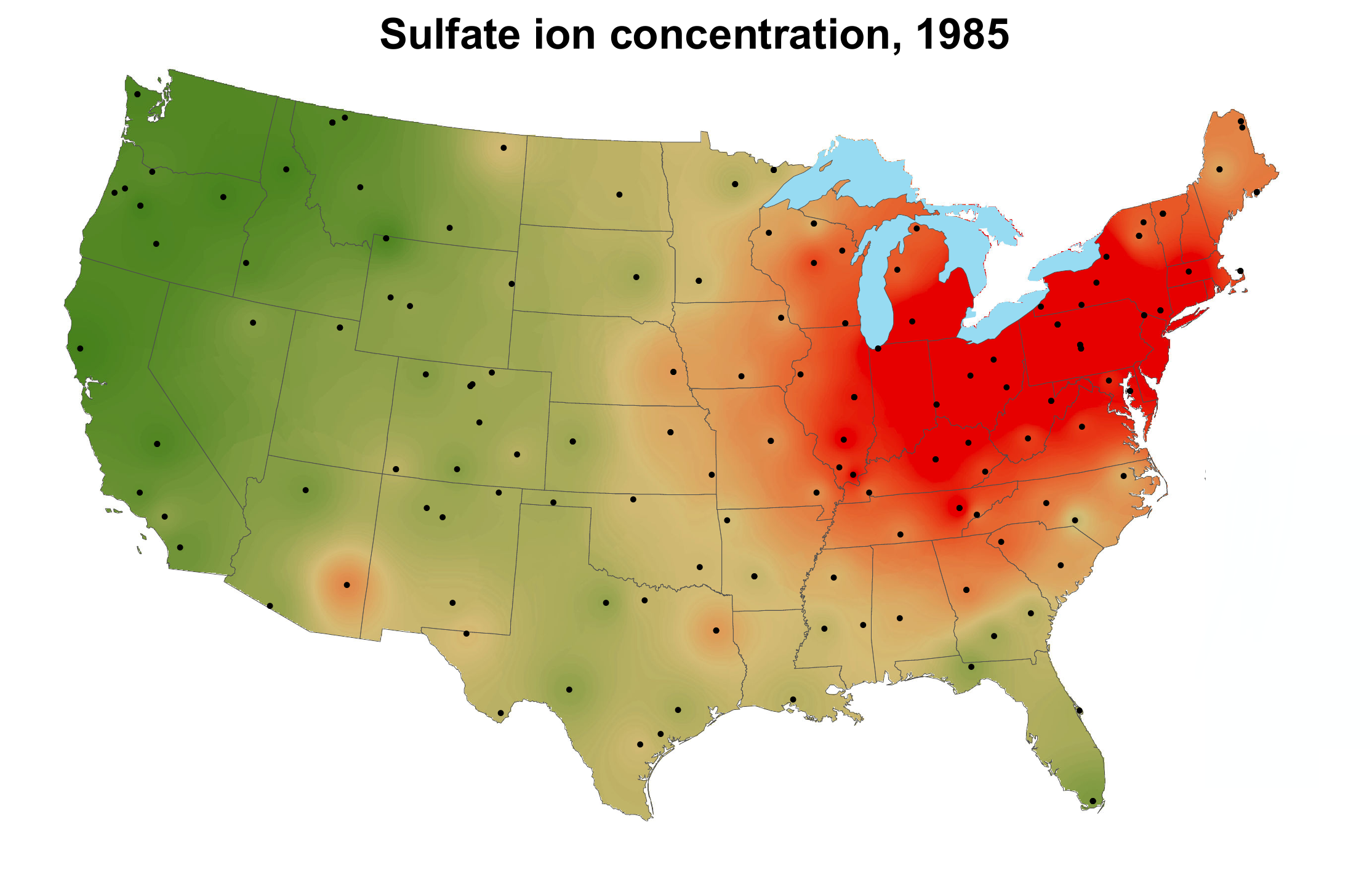} & \includegraphics[width=0.47\textwidth]{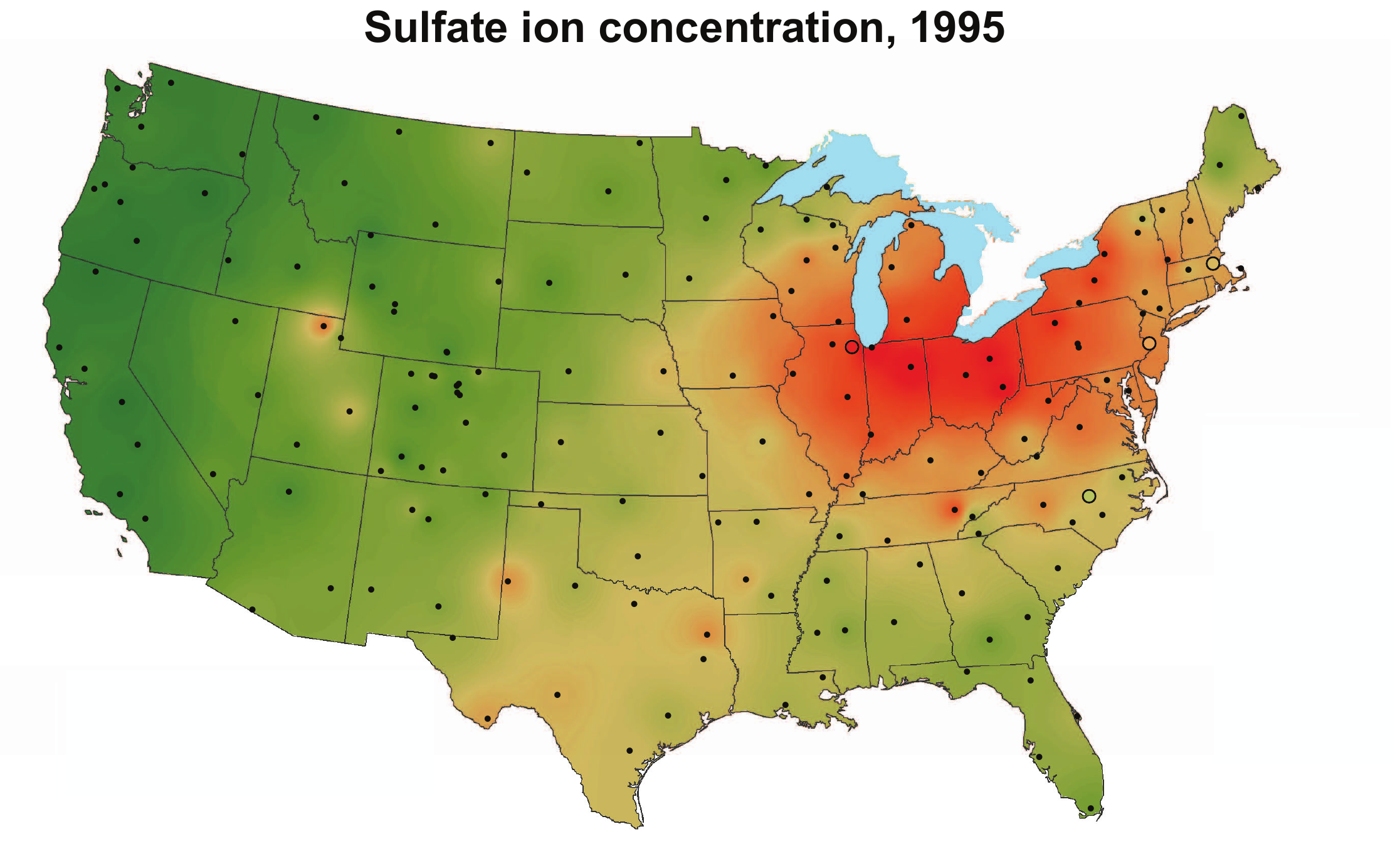} \\
      \includegraphics[width=0.47\textwidth]{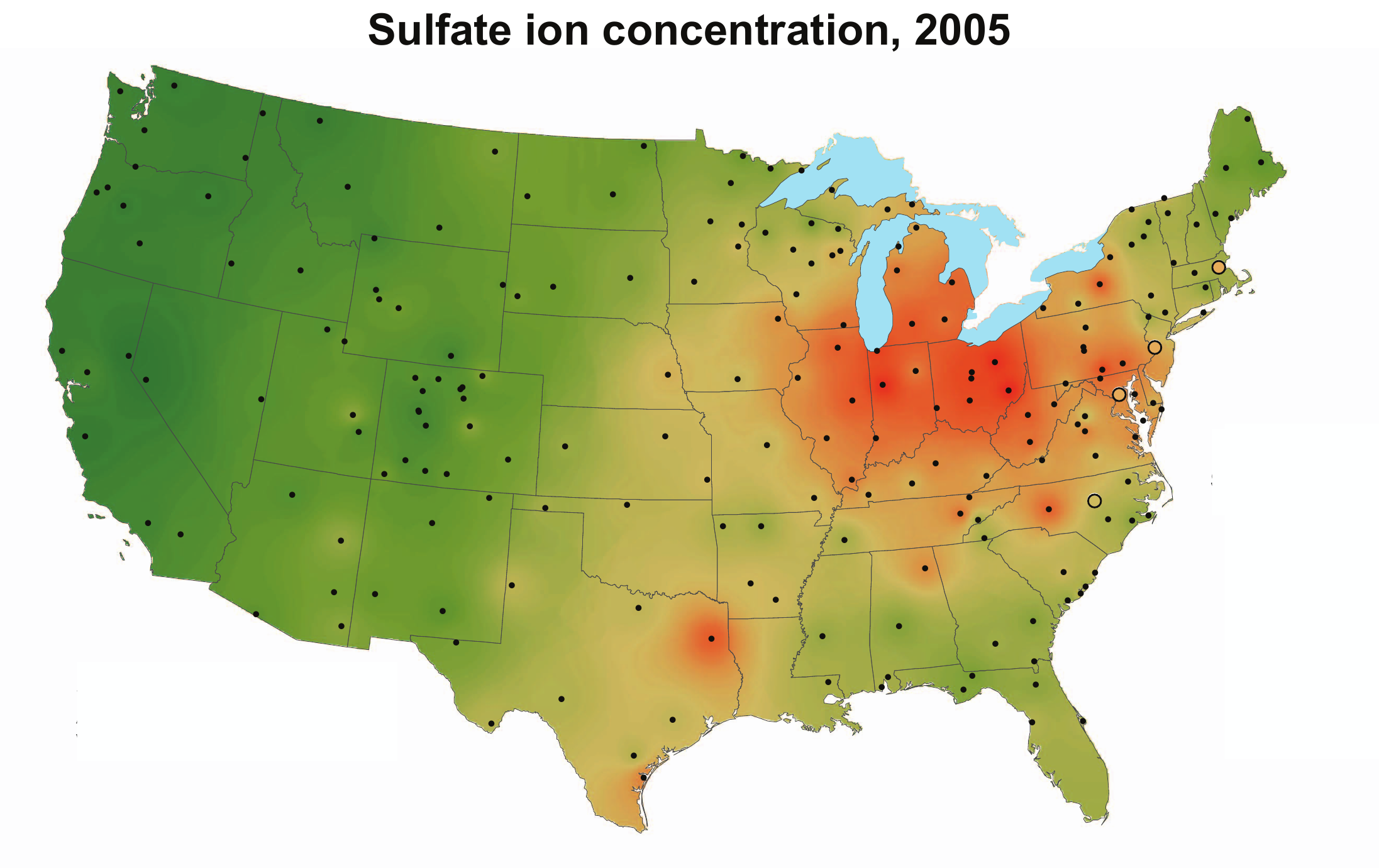} & \includegraphics[width=0.47\textwidth]{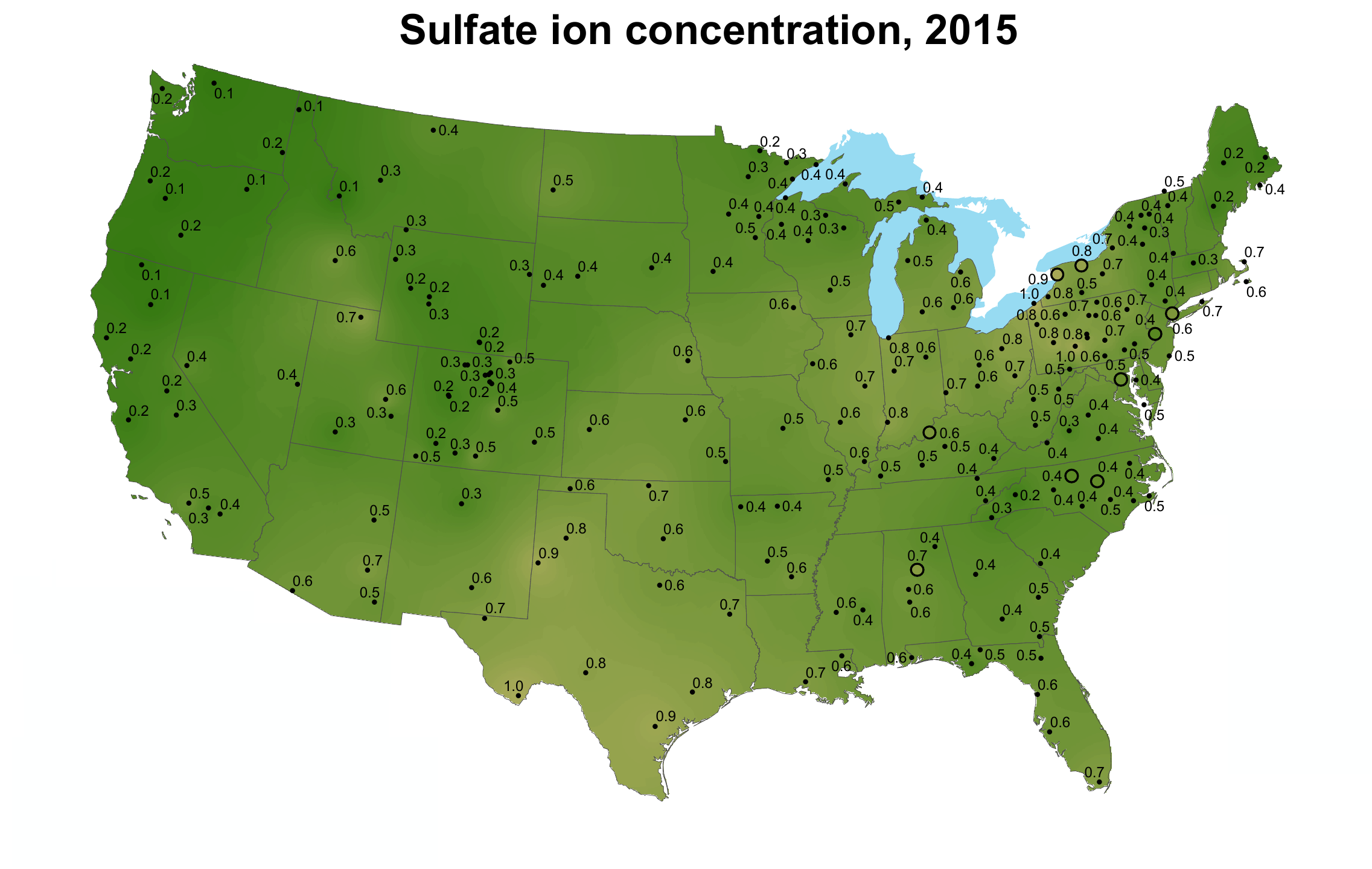} \\[-25pt]
      \multicolumn{2}{c}{\includegraphics[width=0.13\textwidth]{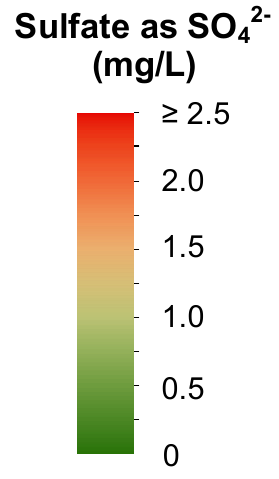}}
\end{tabular}
\caption{Sulfate concentration over time}\label{fig:longtermtrend}
\end{figure}

\begin{figure}[!ht]
\centering
\includegraphics[width = 0.95\textwidth]{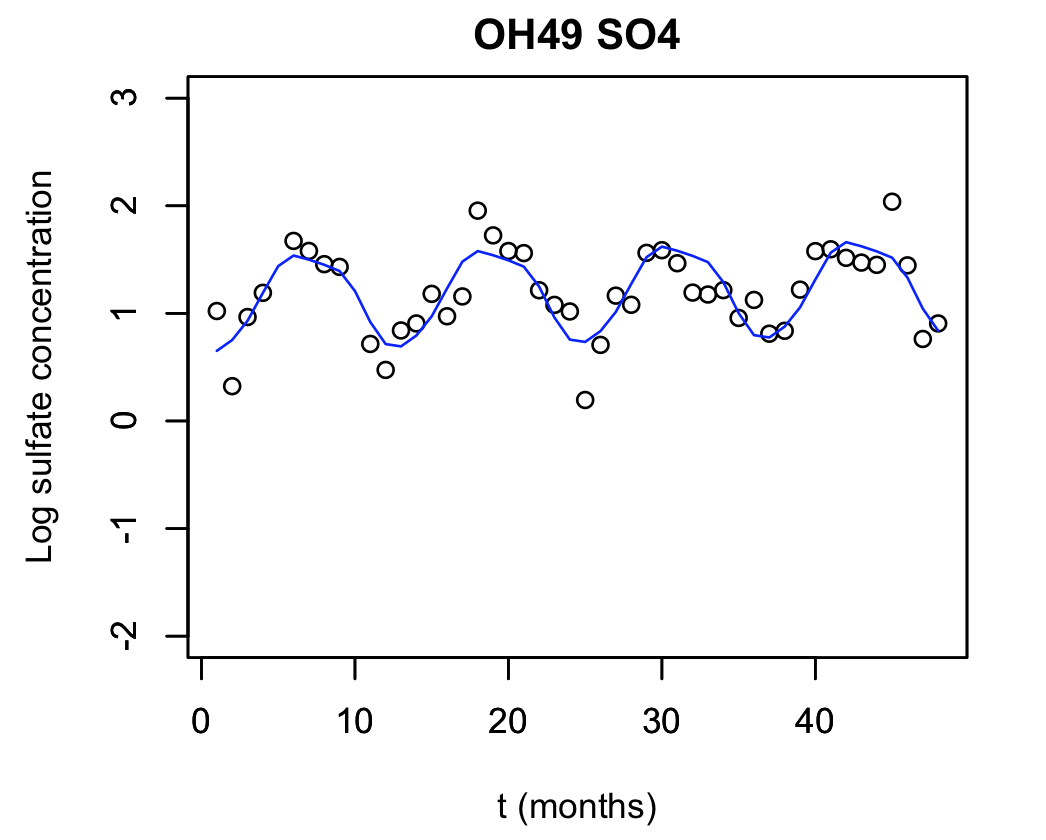}
\caption{Log sulfate concentration over a four-year period at a site}\label{fig:shorttermtrend}
\end{figure}

A quick look at some graphics indicates that there are clear long-term trends; see Figure \ref{fig:longtermtrend}\footnote{Reprinted with the kind permission of the National Atmospheric Deposition Program (NRSP-3). 2019.
NADP Program Office, Wisconsin State Laboratory of Hygiene,
465 Henry Mall, Madison, WI 53706.}, from which we can see that sulfate concentrations are generally trending downward over time. Again, looking at some data, we can easily see periodic trends; see Figure \ref{fig:shorttermtrend}, where we can easily see a yearly periodicity.

The general model that we provide is
%\color{blue} this is more of a question rather than a correction but why
%are we using $\approx$ instead of $=$  \color{black}
\begin{equation} \label{ourmodel}
\reallywidehat{\log(y(t))} = \sum_{i=0}^r \beta_{i}t^i + \sum_{j=1}^{k} \bigg[a_j \cos \bigg(\frac{2\pi jt}{12}\bigg) +  b_j \sin \bigg(\frac{2\pi jt}{12}\bigg)\bigg],
%\widehat{\log(y(t))} = \sum_{i=0}^r \beta_{i}t^i + \sum_{j=1}^{k} \bigg[a_j \cos \bigg(\frac{2\pi jt}{S}\bigg) +  b_j \sin \bigg(\frac{2\pi jt}{S}\bigg)\bigg],
\end{equation}
with the parameters $\beta_i$, $a_i$, and $b_i$ fit by ordinary linear regression. The user can specify the degree $r$ for the polynomial part of the model which we think of as a truncated Taylor series, aimed at capturing aperiodic trends.
Periodic trends are captured via a truncated Fourier series, truncated at level $k$.

We note that \cite{Zidek3} used the following model to de-seasonalize and de-trend the log-transformed monthly sulfate concentration values:
\begin{equation} \label{simplemodel}
%\widehat{\log(y(t))} =  \beta_{1} + \beta_{2}t + a_1\cos\bigg(\frac{2\pi t}{12}\bigg) + b_1\sin\bigg(\frac{2\pi t}{12}\bigg)~.
        \reallywidehat{\log(y(t))} =  \beta_{1} + \beta_{2}t + a_1\cos\bigg(\frac{2\pi t}{12}\bigg) + b_1\sin\bigg(\frac{2\pi t}{12}\bigg)~.
\end{equation}
This is just an affine model $ \beta_{1} + \beta_{2}t$ plus
    a sinusoidal model with monthly periodicity and intercept $a_1$.
    The simple model \eqref{simplemodel} is   \eqref{ourmodel} with $r=1$ and $k=1$.
  We found that \eqref{simplemodel} did well at normalizing the errors for
  certain sites, but some sites, such as  "AL10" and "IN41",  \eqref{simplemodel}
  did not do so well. So rather than
  fix \eqref{simplemodel} as the model for our R package, we provide the flexibility of \eqref{ourmodel}.

  \begin{figure}[t!]
\centering
\begin{tabular}{cc}
  \centering
  \includegraphics[width=.45\linewidth]{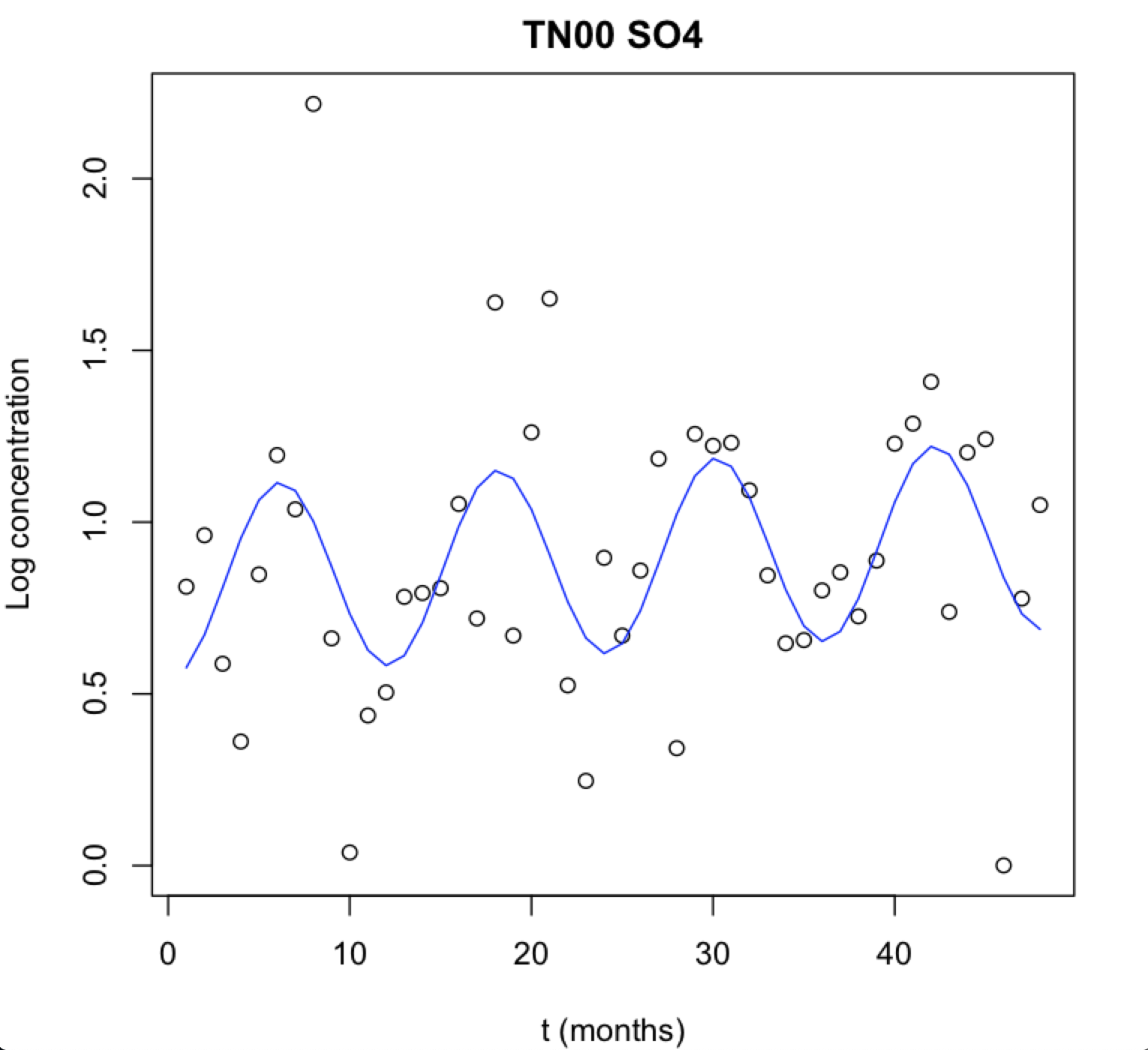}
   &
  \centering
  \includegraphics[width=.45\linewidth]{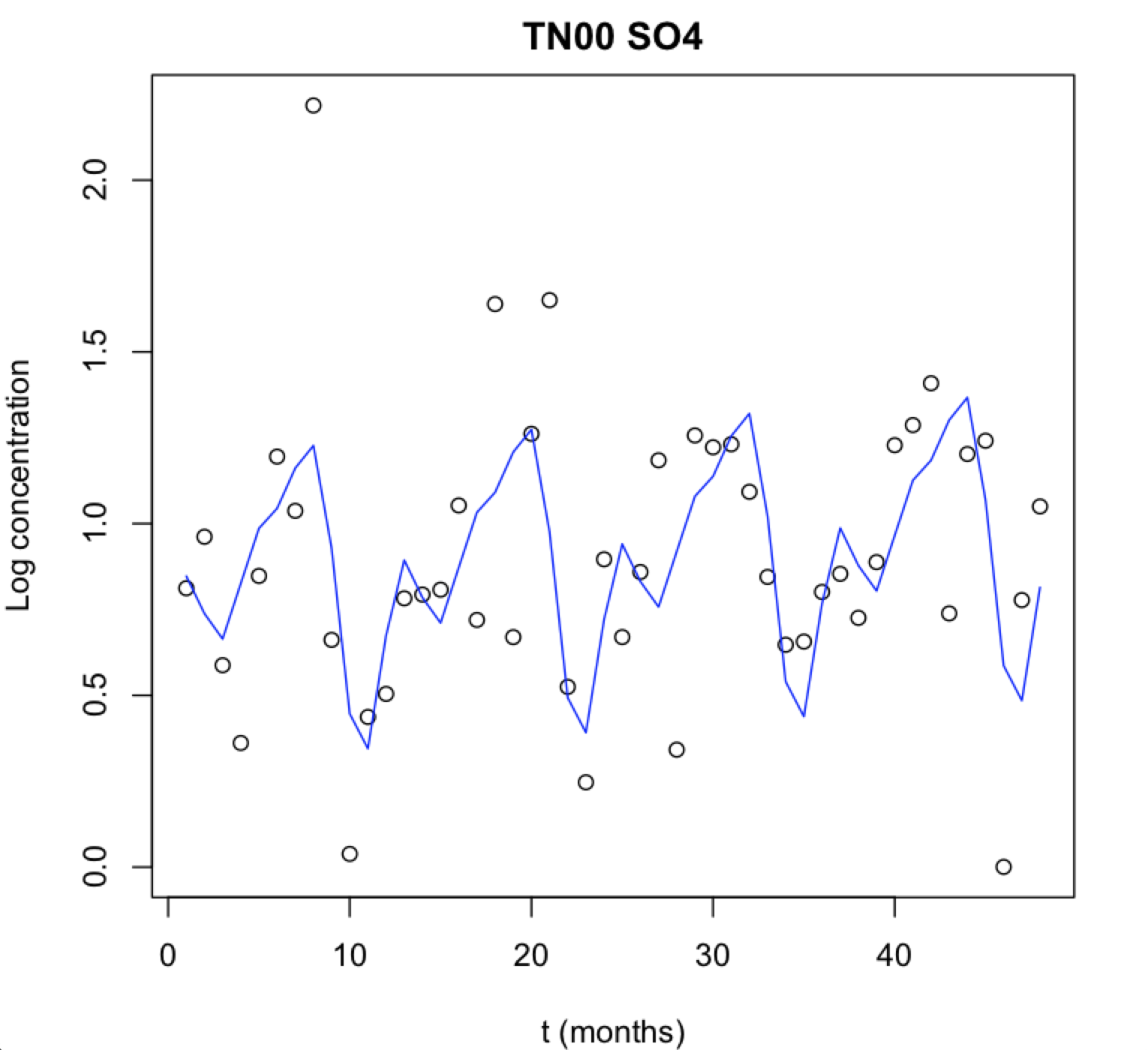}
  %\caption{Plot of fit of eq.1, k = 3, r = 1, S = 12}
\end{tabular}
\caption{Log sulfate concentrations at site "TN00"}
\label{fig:diff}
\end{figure}

In Figure \ref{fig:diff}, we provide an example where \eqref{ourmodel} with $r=1$ and $k=3$ is a much
better fit for the log concentration of sulfate at site "TN00" than \eqref{simplemodel}. The plot on the left shows the fit of model  \eqref{simplemodel}, and the plot on the right shows the fit of our model \eqref{ourmodel}.

 To produce the covariance matrix, we need error values for each time point. Missing values in the NADP/NTN data set means that each set of error values produced by the model may vary in size. So we have filled in missing values for each site and time point by sampling from a normal distribution with the mean being the predicted value by the univariate model and the standard deviation being the standard error of the univariate model.

\section{MESgenCov}\label{sec:Imp}

Our R package \textbf{MESgenCov} can be obtained from
\url{https://github.com/hessakh/MESgenCov} and installed using the \texttt{devtools} library.

\begin{knitrout}
\definecolor{shadecolor}{rgb}{0.969, 0.969, 0.969}\color{fgcolor}\begin{kframe}
\begin{alltt}
 \hlcom{#Install and load MESgenCov}
 \hlkwb{>}   \hlkwd{install_github}\hlstd{(}\hlstr{"hessakh/MESgenCov"}\hlstd{)}
 \hlkwb{>}   \hlkwd{library}\hlstd{(}\hlstr{MESgenCov}\hlstd{)}
\end{alltt}
\end{kframe}
\end{knitrout}

%\textbf{MESgenCov}  contains functions in the \texttt{S3} class to create a covariance matrix from the desired subset of NADP/NTN data. The function \verb;getCov(); returns a covariance matrix, a list of univariate model summaries, and a table of normality tests produced by the MVN R package (see \cite{MVN}). The covariance matrix is produced from a subset of the NADP/NTN data that is specified by the user. For sites with missing data, \verb;getCov(); fills in predicted values based on the univariate model for each site (see \S\ref{sec:3.1}).

\textbf{MESgenCov}  contains functions in the \texttt{S3} class to create a covariance matrix from the desired subset of NADP/NTN data. The function \verb;getCov(); returns a covariance matrix, a list of univariate model summaries, and a table of normality tests produced by the MVN R package (see \cite{MVN}). The multivariate analysis is  used to asses the validity of the covariance matrix to be used as input for the MESP. The user can  make adjustments to the input of \verb;getCov(); so as to obtain a covariance matrix for a multivariate Gaussian vector,
which is then valid for use in the MESP.

 To avoid sites with a small sample size for the specified time-frame, the function \verb;getSites(); outputs a vector of the sites with the largest sample of data for a given time-frame and measured chemical (see \S\ref{sec:3.2}). To find sites that are spatially "spread out" but have at least some specified sample size, the function \verb;maxDistSites(); can be used to obtain a list of geographically sparse sites (see \S\ref{sec:3.2}). Finally,  in the case where the residuals from a univariate model do
not appear to be normally distributed, the function \texttt{lambertWtransform()}
allows the user to transform the  residuals (from a univariate model) using the R package
LambertW: Probabilistic Models to Analyze and Gaussianize
      Heavy-Tailed, Skewed Data (see \cite{LambertWpackage}).
      This can be very effective in situations where the distributions seems to
      have heavy tails and some skewness (see \cite{LambertW2}  and\S\ref{sec:3.3}).

  \subsection{\texttt{getCov}}
  \label{sec:3.1}
  \verb;getCov(); takes a 14 column data frame as input where each column corresponds to one of the user-specifications shown in Figure \ref{fig:param}. The 14 specifications in the input allow the user to specify the subset of data to analyze and gives the user options in displaying different parts of the analysis.
%A table explaining the functions of each column in the input is in Figure \ref{fig:param}.

  %%table
\subsubsection{Input}
%\begin{knitrout}
%\definecolor{shadecolor}{rgb}{0.969, 0.969, 0.969}\color{fgcolor}
%\begin{table}[H]
\begin{figure}[!ht]
\centering
\begin{tabular}{>{\bfseries\leavevmode\color{black}}l>{\raggedright\arraybackslash}p{20em}}
\toprule
Arguments & Definition\\
\midrule
startdateStr & Date and time of when to start analyzing the data, in the format = m/d/y H:M\\
\rowcolor{gray!6}  enddateStr & Date and time of when to stop analyzing the data, in the format = m/d/y H:M\\
comp & String of pollutant or acidity level to be analyzed, the pollutants name should be used as it appears in weeklyConc\\
\rowcolor{gray!6}  use36 & TRUE if default 36 sites should be added, FALSE otherwise\\
\addlinespace
siteAdd & List of strings of siteIDs that should be analyzed\\
\rowcolor{gray!6}  outlierDatesbySite & List of sites where outliers should be analyzed\\
siteOutliers & List of sites where outliers should be removed\\
\rowcolor{gray!6}  removeOutliers & Specify siteID string for outlier analysis\\
plotMulti & TRUE if multivariate analysis plots should be displayed, FALSE otherwise\\
\addlinespace
\rowcolor{gray!6}  sitePlot & Specify list of siteIDs to be plotted\\
plotAll & TRUE if plots for all sites should be displayed, FALSE otherwise\\
\rowcolor{gray!6}  writeMat & TRUE if  .mat file of the resulting covariance matrix should be written in the working directory\\
%seas & Approximate periodicity of data, typically 12 for monthly data\\
\rowcolor{gray!6}  r & Integer <=5, see univariate model\\
k & Integer <= 5, see univariate model\\
\bottomrule
\end{tabular}
\caption{Input parameters for \texttt{getCov()}} \label{fig:param}
\end{figure}
%\end{table}
%\color{blue} I commented p out of the table b/c it's no longer in the model \color{black}
%\end{knitrout}

\FloatBarrier

%\color{green}This part should come after the table, along with it's code chunk that's currently split in two: \color{blue}

A default set of inputs can be found in the stored data frame "defaultInput". Each column of "defaultInput" is an argument in the function \verb;getCov();. After storing "defaultInput" in a variable in the user's workspace, the input can be changed.
For example, below we  store the "defaultInput" data frame in a variable "df", and then change the end date:

\begin{knitrout}
\definecolor{shadecolor}{rgb}{0.969, 0.969, 0.969}\color{fgcolor}\begin{kframe}
\begin{alltt}
\hlcom{#Load defaultInput data frame and store in df}
 \hlkwb{>}   \hlkwd{data}\hlstd{(}\hlstr{"defaultInput"}\hlstd{)}
 \hlkwb{>}   \hlstd{df} \hlkwb{<-} \hlstd{defaultInput}
 \hlkwb{>}   \hlstd{df}
\end{alltt}
\begin{verbatim}
       startdateStr   enddateStr      use36   siteAdd
    1  01/01/83 00:00 12/31/86 00:00  TRUE    NULL
      outlierDatesbySite siteOutliers comp plotMulti sitePlot
    1 NULL               NULL         SO4  FALSE     NULL
      plotAll writeMat r k
      FALSE   FALSE    1 1
\end{verbatim}
\begin{alltt}
\hlcom{#Change the end date to extend the sample of data taken from}
\hlcom{ weeklyConc}
 \hlstd{df}\hlopt{$}\hlstd{enddateStr}   \hlkwb{<-} \hlstr{"12/31/88 00:00"}
\end{alltt}
\end{kframe}
\end{knitrout}

\subsubsection{Output}

The function \verb;getCov(); produces a list with the the following elements:

\begin{tabular}{>{\bfseries\leavevmode\color{black}}l>{\raggedright\arraybackslash}p{20em}}
\toprule
Output & Definition\\
\midrule
\rowcolor{gray!6}  cov & Covariance matrix produced by univariate model residuals\\
listMod & List of univariate model summaries produced by lm()\\
\rowcolor{gray!6}  sites & List of sites that were analyzed\\
mvn & Output of the MVN package\\
\rowcolor{gray!6}  univariateTest & Univariate test output, also by the MVN package\\
\addlinespace
residualData & Data frame of residuals produced by the univariate model\\
\rowcolor{gray!6}  residualDataNA & Data frame of residuals, where missing values are left as NA\\
rosnerTest & Output of the Rosner test for outlier analysis produced by the EnvStats package; see \cite{EnvStats-book}\\
\rowcolor{gray!6}  pred & List of predicted values produced by the univariate model for each site\\
\bottomrule
\end{tabular}
\smallskip

\noindent Next, we demonstrate how to access these elements and certain plots after running the function \verb;getCov();.

\begin{enumerate}
\item Multivariate and univariate normality
\begin{knitrout}
\definecolor{shadecolor}{rgb}{0.969, 0.969, 0.969}\color{fgcolor}\begin{kframe}
\begin{alltt}
\hlcom{#Change part of the input data frame df}
\hlkwb{>}   \hlstd{df}\hlopt{$}\hlstd{plotMulti} \hlkwb{<-} \hlnum{TRUE}
\hlcom{#Change univariate model parameter from 1 to 3 }
\hlkwb{>}   \hlstd{df}\hlopt{$}\hlstd{k} \hlkwb{<-} \hlnum{3}
\hlcom{#Store output in variable g so that the list of outputs given }
\hlcom{ by getCov() can be called}
\hlkwb{>}   \hlstd{g} \hlkwb{<-} \hlkwd{getCov}\hlstd{(df)}
\end{alltt}
\end{kframe}
\end{knitrout}

\begin{center}
\includegraphics[width = 65mm]{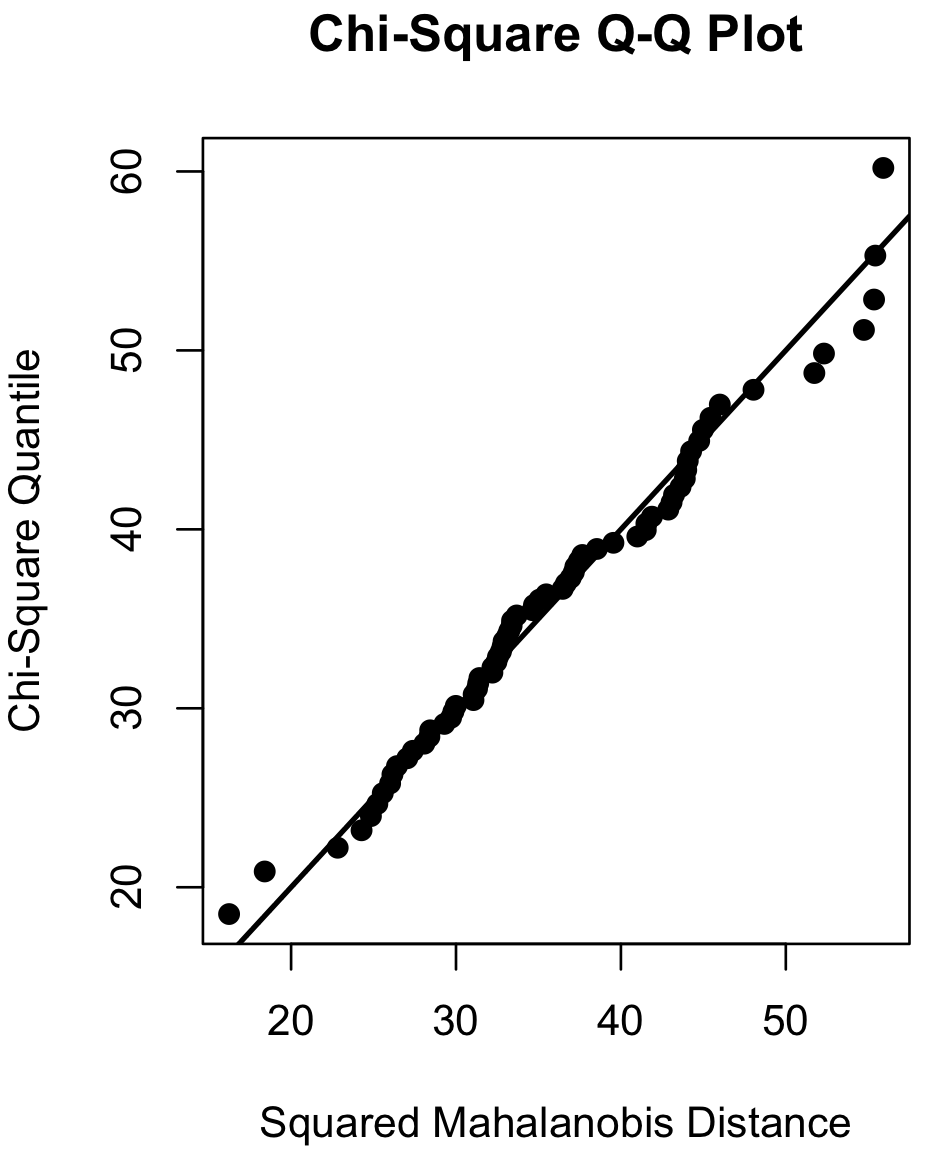}
\end{center}

\begin{knitrout}
\definecolor{shadecolor}{rgb}{0.969, 0.969, 0.969}\color{fgcolor}\begin{kframe}
\begin{alltt}
\hlkwb{>}  \hlstd{g}\hlopt{$}\hlstd{univariateTest}
\begin{verbatim}
           Test  Variable Statistic   p value Normality
 1 Shapiro-Wilk  AL10SO4     0.9347     0.001       NO
 2 Shapiro-Wilk  IL11SO4     0.9894    0.8121      YES
 3 Shapiro-Wilk  IL18SO4     0.9909    0.8854      YES
 4 Shapiro-Wilk  IL19SO4     0.9184     2e-04       NO
 5 Shapiro-Wilk  IL35SO4     0.8709    <0.001       NO
\end{verbatim}
\end{alltt}
\end{kframe}
\end{knitrout}

\item Outlier test for specific tests
\begin{knitrout}
\definecolor{shadecolor}{rgb}{0.969, 0.969, 0.969}\color{fgcolor}\begin{kframe}
\begin{alltt}
  \hlstd{df}\hlopt{$}\hlstd{siteOutliers} \hlkwb{<-} \hlkwd{list}\hlstd{(}\hlkwd{c}\hlstd{(}\hlstr{"IN41"}\hlstd{))}
  \hlstd{df}\hlopt{$}\hlstd{sitePlot} \hlkwb{<-} \hlkwd{list}\hlstd{(}\hlkwd{c}\hlstd{(}\hlstr{"IN41"}\hlstd{))}
  \hlstd{g} \hlkwb{<-} \hlkwd{getCov}\hlstd{(df)}
  \hlstd{i} \hlkwb{<-} \hlkwd{match}\hlstd{(}\hlstr{"IN41"}\hlstd{,g}\hlopt{$}\hlstd{sites)}
  \hlstd{g}\hlopt{$}\hlstd{rosnerTest[[i]]}\hlopt{$}\hlstd{all.stats}
\end{alltt}
\begin{verbatim}
   i  Mean.i   SD.i   Value Obs.Num  R.i+1 lambda.i+1 Outlier
 1 0 -0.0069 0.2814 -0.9359      25 3.3013     3.2680    TRUE
 2 1  0.0062 0.2604 -0.7194      30 2.7862     3.2628   FALSE
 3 2  0.0165 0.2471  0.7215      66 2.8533     3.2576   FALSE
\end{verbatim}
\end{kframe}
\end{knitrout}

By changing the input, we can remove the outliers detected by the Rosner test. Note that the plots are generated after running
\verb;getCov();. Furthermore, \verb;getCov(); does not need to be stored in a variable to generate the plots.
%bottom code should be based on actual outliers, make sure to change accordingly!
\begin{knitrout}
\definecolor{shadecolor}{rgb}{0.969, 0.969, 0.969}\color{fgcolor}\begin{kframe}
\begin{alltt}
\hlcom{#Remove month 25 from site IN41's pollutant concentration data}
\hlkwb{>}  \hlstd{df}\hlopt{$}\hlstd{outlierDatesbySite} \hlkwb{<-} \hlkwd{c}\hlstd{(}\hlstr{"IN41"}\hlstd{,}\hlnum{25}\hlstd{)}
\hlkwb{>}  \hlkwd{getCov}\hlstd{(df)}
\end{alltt}
\end{kframe}
\end{knitrout}
\begin{center}
\includegraphics[width = \textwidth]{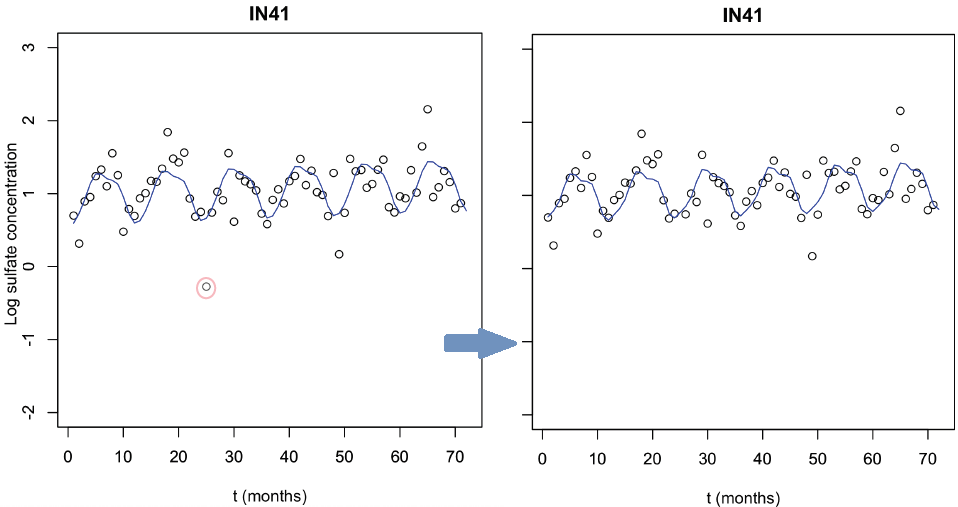}
\end{center}

\item Outlier test for all sites

\begin{knitrout}
\definecolor{shadecolor}{rgb}{0.969, 0.969, 0.969}\color{fgcolor}\begin{kframe}
\begin{alltt}
 \hlcom{#take sites used in analysis in g and run outlier test}
  \hlstd{df}\hlopt{$}\hlstd{siteOutliers} \hlkwb{<-} \hlkwd{list}\hlstd{(g}\hlopt{$}\hlstd{sites)}
 \hlcom{#Remove data points identified as outliers from these sites}
  \hlstd{df}\hlopt{$}\hlstd{removeOutliers} \hlkwb{<-} \hlkwd{list}\hlstd{(g}\hlopt{$}\hlstd{sites)}
  \hlstd{g} \hlkwb{<-} \hlkwd{getCov}\hlstd{(df)}
\end{alltt}
\end{kframe}
\end{knitrout}

\item Plot all sites
\begin{knitrout}
\definecolor{shadecolor}{rgb}{0.969, 0.969, 0.969}\color{fgcolor}\begin{kframe}
\begin{alltt}
\hlkwb{>}  \hlstd{df}\hlopt{$}\hlstd{plotAll} \hlkwb{<-} \hlnum{TRUE}
\hlkwb{>}  \hlkwd{getCov}\hlstd{(df)}
\end{alltt}
\end{kframe}
\end{knitrout}

\begin{center}
  \includegraphics[width=\textwidth]{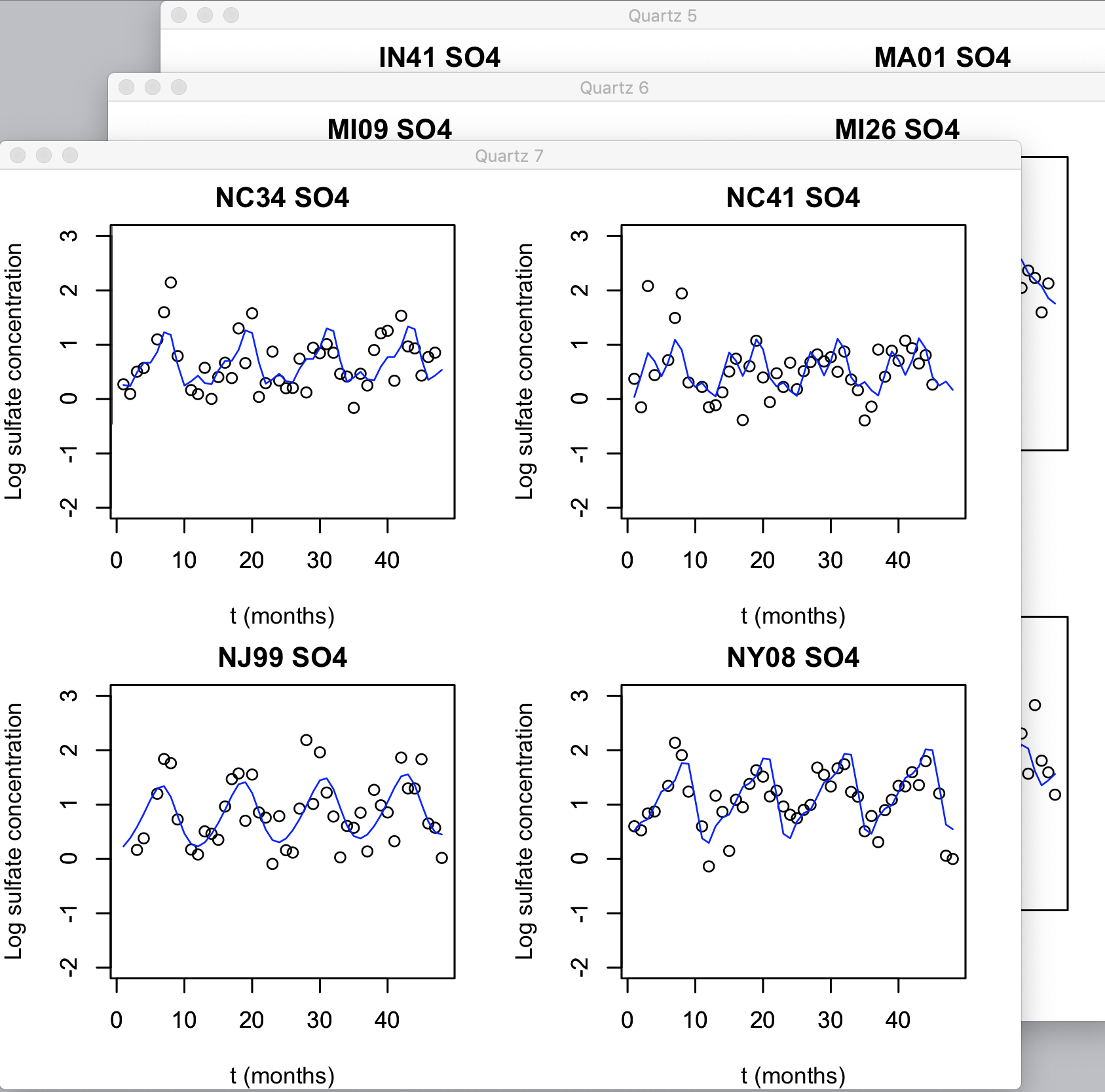}
\end{center}

\item Output all MVN package analysis\\
\noindent
\\
\noindent
The following output is from a call to the MVN package that produces multivariate analysis based on the Mardia method and univariate analysis based on the Shapiro-Wilson method.

\begin{knitrout}
\definecolor{shadecolor}{rgb}{0.969, 0.969, 0.969}\color{fgcolor}\begin{kframe}
\begin{alltt}
\hlcom{#Remove default list of sites so that their data is not }
\hlcom{ analyzed}
\hlkwb{>}  \hlstd{df}\hlopt{$}\hlstd{use36} \hlkwb{<-} \hlnum{FALSE}
\hlcom{#Add new site list}
\hlkwb{>}  \hlstd{df}\hlopt{$}\hlstd{siteAdd} \hlkwb{<-} \hlkwd{list}\hlstd{(}\hlkwd{c}\hlstd{(}\hlstr{"NY52"}\hlstd{,} \hlstr{"TN11"}\hlstd{,} \hlstr{"IL63"}\hlstd{))}
\hlcom{#Remove any set of sites and pollutant combinations}
\hlcom{ that had been previously added}
\hlkwb{>}  \hlstd{df}\hlopt{$}\hlstd{siteOutliers} \hlkwb{<-} \hlkwa{NULL}
\hlkwb{>}  \hlstd{df}\hlopt{$}\hlstd{outlierDatesbySite} \hlkwb{<-} \hlkwa{NULL}
\hlkwb{>}  \hlstd{df}\hlopt{$}\hlstd{removeOutliers} \hlkwb{<-} \hlkwa{NULL}
\hlkwb{>}  \hlstd{g} \hlkwb{<-} \hlkwd{getCov}\hlstd{(df)}
\hlcom{#Display full output of the MVN package}
\hlkwb{>}  \hlstd{g}\hlopt{$}\hlstd{mvn}
\end{alltt}
\begin{verbatim}
 $multivariateNormality
              Test Statistic p value Result
 1 Mardia Skewness    14.019  0.1721    YES
 2 Mardia Kurtosis   -0.1465  0.8835    YES
 3             MVN      <NA>    <NA>    YES

 $univariateNormality
           Test  Variable Statistic   p value Normality
 1 Shapiro-Wilk  NY52SO4     0.9821    0.3981    YES
 2 Shapiro-Wilk  TN11SO4     0.9830    0.4385    YES
 3 Shapiro-Wilk  IL63SO4     0.9873    0.6858    YES

 $Descriptives
          n     Mean   Std.Dev  Median     Min    Max
 NY52SO4 72  1.1709e-17 0.2812  0.0333 -0.8815 0.6170
 TN11SO4 72 -1.4991e-02 0.4666 -0.0020 -0.9826 1.4243
 IL63SO4 72  4.8127e-18 0.3016 -0.0373 -0.7353 0.7899
               25th      75th       Skew   Kurtosis
 NY52SO4    -0.2211    0.1840    -0.3702     0.0894
 TN11SO4     -0.326    0.2130     0.4514     0.2981
 IL63SO4   -0.18369    0.1724     0.3287     0.0609
\end{verbatim}
\end{kframe}
\end{knitrout}

The specific call of the MVN package is

\begin{knitrout}
\definecolor{shadecolor}{rgb}{0.969, 0.969, 0.969}\color{fgcolor}\begin{kframe}
\begin{alltt}
\hlkwb{>}  \hlkwd{mvn}(dfRes[,-1], subset = NULL, mvnTest = \hlstr{"mardia"},
 covariance = TRUE, tol = 1e-25, alpha = 0.5, scale = FALSE,
 desc = TRUE, transform = \hlstr{"none"},univariateTest = \hlstr{"SW"},
 univariatePlot = \hlstr{"none"}, multivariatePlot =\hlstr{"none"},
 multivariateOutlierMethod = \hlstr{"none"}, bc = FALSE, bcType =
 \hlstr{"rounded"}, showOutliers = FALSE, showNewData = FALSE).
\end{alltt}
\end{kframe}
\end{knitrout}

See \cite{MVN} for details on the MVN package.

\item Covariance matrix
%%%%Parsing results
\begin{knitrout}
\definecolor{shadecolor}{rgb}{0.969, 0.969, 0.969}\color{fgcolor}\begin{kframe}
\begin{alltt}
\hlkwb{>}  \hlstd{g} \hlkwb{<-} \hlkwd{getCov}\hlstd{(df)}
\hlcom{#Print covariance matrix}
\hlkwb{>}  \hlkwd{round}\hlstd{(g}\hlopt{$}\hlstd{cov,}\hlkwc{digits} \hlstd{=} \hlnum{4}\hlstd{)}
\end{alltt}
\begin{verbatim}
         NY52SO4 TN11SO4 IL63SO4
 NY52SO4  0.0791  0.0047  0.0009
 TN11SO4  0.0047  0.2177  0.0185
 IL63SO4  0.0009  0.0185  0.0909
\end{verbatim}
\end{kframe}
\end{knitrout}

\item Save covariance matrix as a .mat file (to populate an instance of the MESP, for example).

This is done by simply setting the
%12th column in
input data frame
attribute
 \texttt{writeMat} to TRUE. The .mat file will be saved to the user's current working directory as covSites.mat. For processing further with Matlab, use the (Matlab) `load' command.

\begin{knitrout}
\definecolor{shadecolor}{rgb}{0.969, 0.969, 0.969}\color{fgcolor}\begin{kframe}
\begin{alltt}
\hlcom{#Write cov into .mat file in current directory}
\hlkwb{>}  \hlstd{df}\hlopt{$}\hlstd{writeMat} \hlkwb{<-} \hlnum{TRUE}
\hlkwb{>}  \hlstd{g}           \hlkwb{<-} \hlkwd{getCov}\hlstd{(df)}
\end{alltt}
\end{kframe}
\end{knitrout}

In the case that the user has already generated an output by the function \verb;getCov();, it is possible to also create the .mat file in the following manner.

\begin{knitrout}
\definecolor{shadecolor}{rgb}{0.969, 0.969, 0.969}\color{fgcolor}\begin{kframe}
\begin{alltt}
\hlkwb{>}  \hlkwd{library}\hlstd{(rmatio)}
\hlkwb{>}  \hlkwd{write.mat}\hlstd{(g}\hlopt{$}\hlstd{cov,}\hlkwc{filename} \hlstd{=} \hlstr{"covariance1.mat"}\hlstd{)}
\end{alltt}
\end{kframe}
\end{knitrout}

\item Univariate model summaries
%%%Parsing results
    \begin{knitrout}
    \definecolor{shadecolor}{rgb}{0.969, 0.969, 0.969}\color{fgcolor}
  \begin{kframe}
    \begin{alltt}
\hlkwb{>}  \hlstd{result} \hlkwb{<-} \hlkwd{getCov}\hlstd{(df)}
\hlcom{#Store site list in sites variable}
\hlkwb{>}  \hlstd{sites}  \hlkwb{<-} \hlstd{result}\hlopt{$}\hlstd{sites}
\hlcom{#Find site OH71 index in the list}
\hlkwb{>}  \hlstd{i} \hlkwb{=} \hlkwd{match}\hlstd{(}\hlkwd{c}\hlstd{(}\hlstr{"NY52"}\hlstd{),sites)}
\hlcom{#Use site index to find model summary for NY52}
\hlkwb{>}  \hlstd{result}\hlopt{$}\hlstd{listMod[i]}
    \end{alltt}
    \begin{verbatim}
    [[1]]
    Call:
    lm(formula = y1 ~ I(cos(t*(2*pi/s))) + I(sin(t*(2 *
        pi/s))) + I(cos(t*(2*pi/s)*2)) + I(sin(t*
        (2*pi/s)*2)) + I(cos(t*(2*pi/s)*3)) +
        I(sin(t*(2*pi/s)*3)) + I(t), data = df)

    Residuals:
        Min      1Q  Median      3Q     Max
    -0.5236 -0.1677 -0.0189  0.1818  0.9087

    Coefficients:
                        Estimate Std. Error t value Pr(>|t|)
    (Intercept)           1.1110    0.0712   14.40  < 2e-16***
    I(cos(t*(2*pi/s)))   -0.3541    0.0494   -9.75  2.8e-14***
    I(sin(t*(2*pi/s)))   -0.1109    0.0498   -2.55    0.013*
    I(cos(t*(2*pi/s)*2))  0.0500    0.0494   -0.37    0.716
    I(sin(t*(2*pi/s)*2))  0.0797    0.0494    2.13    0.037*
    I(cos(t*(2*pi/s)*3))  0.0391    0.0494   -0.20    0.845
    I(sin(t*(2*pi/s)*3))  0.0536    0.0494   -0.52    0.604
    I(t)                 -0.0010    0.0017   -0.61    0.546
    ---
 Signif. codes: 0 '***' 0.001 '**' 0.01 '*' 0.05 '.' 0.1 ' ' 1

    Residual standard error: 0.2961 on 64 degrees of freedom
    Multiple R-squared:  0.6265,	Adjusted R-squared:  0.5857
    F-statistic: 15.34 on 7 and 64 DF,  p-value: 1.327e-11
    \end{verbatim}
    \end{kframe}
    \end{knitrout}

\item Output data frame of residuals\\
\begin{knitrout}
\definecolor{shadecolor}{rgb}{0.969, 0.969, 0.969}\color{fgcolor}\begin{kframe}
\begin{alltt}
\hlkwb{>}   \hlstd{g} \hlkwb{<-} \hlkwd{getCov}\hlstd{(df)}
\hlcom{#Display dataframe containing the residuals}
\hlcom{ from the fitted univariate model}
\hlkwb{>}    \hlstd{g}\hlopt{$}\hlstd{residualData[}\hlnum{1}\hlopt{:}\hlnum{5}\hlstd{,]}
\end{alltt}
\begin{verbatim}
      NY52SO4     TN11SO4    IL63SO4
 1  0.1959813 -0.44377735  0.2824803
 2  0.6170340 -0.38791938 -0.3495080
 3 -0.4510128  1.05519620 -0.1158200
 4 -0.1580145 -0.01789642 -0.1297776
 5 -0.3218159 -0.47833685 -0.3415119
\end{verbatim}
\end{kframe}
\end{knitrout}

\end{enumerate}

  \subsection{Functions for getting a vector of sites}
    \label{sec:3.2}
\verb;getCov(); takes site lists as input. The function \texttt{getSites()} produces a list of sites with available data for a specified time frame. The code below produces a list of 36 sites with the most weekly data between the years 1983--1986.

\begin{knitrout}
\definecolor{shadecolor}{rgb}{0.969, 0.969, 0.969}\color{fgcolor}\begin{kframe}
\begin{alltt}
\hlkwb{>}  \hlstd{result} \hlkwb{<-} \hlkwd{getSites}\hlstd{(}\hlstr{"01/01/83 00:00"}\hlstd{,}\hlstr{"12/31/86 00:00"}\hlstd{,}\hlnum{36}\hlstd{,}\hlnum{104}\hlstd{,}
             \hlstr{"SO4"}\hlstd{,}\hlstr{""}\hlstd{)}
\hlkwb{>}  \hlstd{result}\hlopt{$}\hlstd{finalList}
\end{alltt}
\begin{verbatim}
   [1] "OH71" "NY08" "WV18" "MI53" "NH02" "OH49" "PA42" "ME09"
   [9] "IN34" "MA13" "NY52" "NY10" "WA14" "NY20" "OH17" "ME00"
  [17] "TN00" "IL63" "MI99" "WI28" "IN41" "PA29" "WI36" "ME02"
  [25] "MI09" "MO05" "NC03" "NJ99" "PA15" "CO19" "MN18" "WI37"
  [33] "AR27" "KS31" "ME98" "MO03"
\end{verbatim}
\end{kframe}
\end{knitrout}

The 4th input specifies the minimum sample of weekly data required to be included in the produced list, and the last input tells the function to only look at sites in the Northern region of the US. The options for region are "W", "S", and "N"; see \hyperref[appB]{Appendix B} for the geographic split.

\begin{knitrout}
\definecolor{shadecolor}{rgb}{0.969, 0.969, 0.969}\color{fgcolor}\begin{kframe}
\begin{alltt}
\hlkwb{>}  \hlstd{NSites} \hlkwb{<-} \hlkwd{getSites}\hlstd{(}\hlstr{"01/01/83 00:00"}\hlstd{,}\hlstr{"12/31/86 00:00"}\hlstd{,}\hlnum{36}\hlstd{,}\hlnum{104}\hlstd{,}
             \hlstr{"SO4"}\hlstd{,}\hlstr{"N"}\hlstd{)}
\hlkwb{>}  \hlstd{NSites}\hlopt{$}\hlstd{finalList}
\end{alltt}
\begin{verbatim}
   [1] "OH71" "NY08" "MI53" "NH02" "OH49" "PA42" "ME09" "IN34"
   [9] "MA13" "NY52" "NY10" "NY20" "OH17" "ME00" "MI99" "WI28"
  [17] "IN41" "PA29" "WI36" "ME02" "MI09" "NJ99" "PA15" "MN18"
  [25] "WI37" "ME98" "IL11" "IL18" "MN16" "MI26" "NE15" "VT01"
  [33] "NY99" "MA01" "MA08" "MN27"
\end{verbatim}
\end{kframe}
\end{knitrout}

The function \texttt{maxDistSites()} prioritizes sites that are farther away from each other. This function takes the same arguments as input as \texttt{getSites()}, except for the last argument where instead of specifying a region, the user can specify which site should be included first. If  the last argument is 1, then the site with the greatest amount of data for the specified time period will be chosen; if the last argument is 2, then the site with the second greatest amount of data will be chosen;
etc.
%\color{blue} output was missing \color{black}
\begin{knitrout}
\definecolor{shadecolor}{rgb}{0.969, 0.969, 0.969}\color{fgcolor}\begin{kframe}
\begin{alltt}
\hlkwb{>}   \hlstd{maxdist} \hlkwb{<-} \hlkwd{maxDistSites}\hlstd{(}\hlstr{"01/01/83 00:00"}\hlstd{,}\hlstr{"12/31/86 00:00"}\hlstd{,}\hlnum{36}\hlstd{,}
               \hlnum{104}\hlstd{,}\hlstr{"SO4"}\hlstd{,}\hlnum{1}\hlstd{)}
\hlkwb{>}   \hlstd{maxdist}\hlopt{$}\hlstd{finalList}
\end{alltt}
\begin{verbatim}
   [1] "OH71" "WA14" "TX04" "FL11" "ME00" "WY06" "MN27" "LA12"
   [9] "CA45" "OK00" "NY99" "GA41" "MI99" "AZ03" "MT05" "NC35"
  [17] "MO05" "CO00" "WY99" "IN34" "KY03" "MI09" "FL03" "MA01"
  [25]  "OR10" "PA42" "AR27" "MN16" "TX21" "VT99" "NE15" "VA13"
  [33]  "CO15" "CO22" "NY52" "AR02"
\end{verbatim}
\end{kframe}
\end{knitrout}

 \subsection{Lambert W transformation on univariate data}
    \label{sec:3.3}
  For a number of sites, the residuals produced by our univariate model have skewed distributions with heavy tails. In particular, this is the case for many sites when the sample of data is taken over a period greater than 4 years. To deal with this issue, we have incorporated functions from the LambertW package (see \cite{LambertWpackage}) in the function \texttt{lambertWtransform()} that will allow a user to transform the residuals produced by the deterministic univariate model. The LambertW package estimates the parameters that fit a Lambert W distribution on the given univariate data. Then the underlying Gaussian distribution implied by the Lambert W distribution is extracted and is used for the multivariate analysis in the function \texttt{lambertWtransform()}. The \texttt{lambertWtransform()} function takes the following as input: a data frame of residuals, and two logical inputs specifying whether to plot the multivariate qq plot and whether to produce the .mat file containing the covariance matrix with the Lambert W transformed residuals. Details on the algorithms that perform the transformation can be found in \cite{LambertW2}. Here we provide an example where we transform the residuals of 50 sites stored in an internal dataset, named "dfRes50".

\begin{knitrout}
\definecolor{shadecolor}{rgb}{0.969, 0.969, 0.969}\color{fgcolor}\begin{kframe}
\begin{alltt}
\hlkwb{>}  \hlkwd{data}\hlstd{(}\hlstr{"dfRes50"}\hlstd{)}
\hlkwb{>}  \hlstd{loutput} \hlkwb{<-} \hlkwd{lambertWtransform}\hlstd{(}\hlkwc{dfRes}\hlstd{=dfRes50,} \hlkwc{plotMulti}\hlstd{=}\hlnum{FALSE}\hlstd{,}
              \hlkwc{writeMat}\hlstd{=}\hlnum{FALSE}\hlstd{)}
\hlkwb{>}  \hlstd{loutput}\hlopt{$}\hlstd{mvn}\hlopt{$}\hlstd{multivariateNormality}
\end{alltt}
\begin{verbatim}
               Test  Statistic   p value    Result
  1 Mardia Skewness      22800    0.0004        NO
  2 Mardia Kurtosis      0.418    0.6763       YES
  3             MVN       <NA>      <NA>        NO
\end{verbatim}
\end{kframe}
\end{knitrout}

 This function produces a list of four outputs:

  \begin{enumerate}
   \item \texttt{loutput\$mvn} contains the results of applying the multivariate analysis by the MVN package
   \item \texttt{loutput\$cov} contains the covariance matrix produced by the transformed residuals
   \item \texttt{loutput\$newResiduals} contains the data frame of Lambert W transformed         residuals
   \item \texttt{loutput\$univariateTest} contains the univariate tests produced by the MVN function for the   transformed residuals
 \end{enumerate}

\begin{knitrout}
\definecolor{shadecolor}{rgb}{0.969, 0.969, 0.969}\color{fgcolor}\begin{kframe}
\begin{alltt}
\hlkwb{>}  \hlkwd{data}\hlstd{(}\hlstr{"dfRes50"}\hlstd{)}
\hlkwb{>}  \hlstd{dfRes50} \hlkwb{<-} \hlstd{dfRes50}
\hlkwb{>}  \hlstd{loutput} \hlkwb{<-} \hlkwd{lambertWtransform}\hlstd{(}\hlkwc{dfRes}\hlstd{=dfRes50,} \hlkwc{plotMulti}\hlstd{=}\hlnum{FALSE}\hlstd{,}
              \hlkwc{writeMat}\hlstd{=}\hlnum{FALSE}\hlstd{)}
\hlkwb{>}  \hlstd{loutput}\hlopt{$}\hlstd{mvn}
\hlkwb{>}  \hlstd{loutput}\hlopt{$}\hlstd{cov}
\hlkwb{>}  \hlstd{loutput}\hlopt{$}\hlstd{newResiduals}
\hlkwb{>}  \hlstd{loutput}\hlopt{$}\hlstd{univariateTest}
\end{alltt}
\end{kframe}
\end{knitrout}

Next, we present an example where we use \texttt{maxDistSites()} to get a list of 50 sites that is geographically sparse and has at least 200 weeks of data between 1986 and 1994. From this list of sites, a covariance and its corresponding multivariate normality test is generated and compared to the Lambert W transformed output.

\begin{knitrout}
\definecolor{shadecolor}{rgb}{0.969, 0.969, 0.969}\color{fgcolor}\begin{kframe}
\begin{alltt}
\hlcom{#Get list of sites}
\hlkwb{>}  \hlstd{maxd}  \hlkwb{<-} \hlkwd{maxDistSites}\hlstd{(}\hlstr{"01/01/86 00:00"}\hlstd{,}\hlstr{"12/31/94 00:00"}\hlstd{,}\hlnum{50}\hlstd{,}
            \hlnum{200}\hlstd{,}\hlstr{"SO4"}\hlstd{,}\hlnum{1}\hlstd{)}
\hlcom{#Create input data frame}
\hlkwb{>}  \hlstd{df} \hlkwb{<-} \hlstd{defaultInput}
\hlcom{#Use list of sites and specification in maxd}
\hlkwb{>}  \hlstd{df}\hlopt{$}\hlstd{siteAdd} \hlkwb{<-} \hlkwd{list}\hlstd{(maxd}\hlopt{$}\hlstd{finalList)}
\hlkwb{>}  \hlstd{df}\hlopt{$}\hlstd{startdateStr} \hlkwb{<-} \hlstd{maxd}\hlopt{$}\hlstd{startDate}
\hlkwb{>}  \hlstd{df}\hlopt{$}\hlstd{use36} \hlkwb{<-} \hlnum{FALSE}
\hlkwb{>}  \hlstd{df}\hlopt{$}\hlstd{comp} \hlkwb{<-} \hlstd{maxd}\hlopt{$}\hlstd{comp}
\hlkwb{>}  \hlstd{df}\hlopt{$}\hlstd{enddateStr} \hlkwb{<-} \hlstd{maxd}\hlopt{$}\hlstd{endDate}
\hlkwb{>}  \hlstd{df}\hlopt{$}\hlstd{writeMat} \hlkwb{<-} \hlnum{TRUE}
\hlkwb{>}  \hlstd{output} \hlkwb{<-} \hlkwd{getCov}\hlstd{(df)}
\hlkwb{>}  \hlstd{output}\hlopt{$}\hlstd{mvn}\hlopt{$}\hlstd{multivariateNormality}
\end{alltt}
\begin{verbatim}
               Test   Statistic  p value  Result
  1 Mardia Skewness       22962  2.6e-05      NO
  2 Mardia Kurtosis      0.2408   0.8097     YES
  3             MVN        <NA>     <NA>      NO
\end{verbatim}
\begin{alltt}
\hlstd{loutput} \hlkwb{<-} \hlkwd{lambertWtransform}\hlstd{(g}\hlopt{$}\hlstd{residualDataNA,} \hlnum{TRUE}\hlstd{,}\hlnum{FALSE}\hlstd{)}
\hlstd{loutput}\hlopt{$}\hlstd{mvn}\hlopt{$}\hlstd{multivariateNormality}
\end{alltt}
\begin{verbatim}
              Test   Statistic   p value  Result
 1 Mardia Skewness       22263    0.2174     YES
 2 Mardia Kurtosis     -1.6039    0.1087     YES
 3             MVN        <NA>     <NA>      YES
\end{verbatim}
\end{kframe}
\end{knitrout}

  \subsection{Internal datasets and their properties}
  \label{sec:3.4}

  We provide five internal covariance matrices that we have produced from geographically sparse lists of sites.
  The exact specifications used to produce the sites can be found in
   \hyperref[appC]{Appendix C}. We offer these covariance matrices for the convenience of the user. We note that although the list of sites are quite spread out geographically, they are not independent. We test the independence of the covariance matrices using a likelihood ratio test (see \cite[p. 275]{rencher12})
%  where they calculate a test statistic $u$ and they compare it to the $\alpha$ level of $\chi^2_{m(m-1)/2}$.
   The test statistic  is
\[
 u := - \left(\nu - \dfrac{2m+5}{6}\right)\log(\det(R)),
 \]
 where $m := $ number of sites, $\nu := m(m+1)/2$, and $R$ is the sample correlation matrix. The null hypothesis $H_0$ is that the variates are independent, and we reject $H_0$ if $u > \chi^2_{m(m-1)/2,\alpha}$ where for our analysis $\alpha = 0.05$. The five covariance matrices are named "maxd1Cov", "maxd2Cov", "maxd3Cov", "maxd4Cov", and "maxd5Cov", and the
  corresponding test statistics $u$ are 21605, 25740, 26928, 25870, and 25137,
  all comfortably giving evidence to reject $H_0$ at the $\alpha = 0.05$ level
  (we reject when $u> 1308$).

 We have made a function available that performs this independence test. Here we indicate how it can be used on a data frame of residuals produced by \verb;getCov();.

\begin{knitrout}
\definecolor{shadecolor}{rgb}{0.969, 0.969, 0.969}\color{fgcolor}\begin{kframe}
\begin{alltt}
\hlkwb{>} \hlstd{maxd1}  \hlkwb{<-} \hlkwd{maxDistSites}\hlstd{(}\hlstr{"01/01/86 00:00"}\hlstd{,}\hlstr{"12/31/94 00:00"}\hlstd{,}\hlnum{50}\hlstd{,}
            \hlnum{250}\hlstd{,}\hlstr{"SO4"}\hlstd{,}\hlnum{1}\hlstd{)}
\hlkwb{>} \hlstd{df}\hlopt{$}\hlstd{comp} \hlkwb{<-} \hlstd{maxd1}\hlopt{$}\hlstd{comp}
\hlkwb{>} \hlstd{df}\hlopt{$}\hlstd{enddateStr} \hlkwb{<-} \hlstd{maxd1}\hlopt{$}\hlstd{endDate}
\hlkwb{>} \hlstd{df}\hlopt{$}\hlstd{startdateStr} \hlkwb{<-} \hlstd{maxd1}\hlopt{$}\hlstd{startDate}
\hlkwb{>} \hlstd{df}\hlopt{$}\hlstd{siteAdd} \hlkwb{<-} \hlkwd{list}\hlstd{(maxd1}\hlopt{$}\hlstd{finalList)}
\hlkwb{>} \hlstd{result}  \hlkwb{<-} \hlkwd{getCov}\hlstd{(df)}
\hlkwb{>} \hlstd{indp}  \hlkwb{<-} \hlkwd{independenceTest}\hlstd{(result}\hlopt{$}\hlstd{residualData)}
\hlkwb{>} \hlstd{indp}\hlopt{$}\hlstd{test}
\end{alltt}
\begin{verbatim}
    chisq dist likelihood ratio   tchisq  independent
  1                       21605  1307.54        FALSE
\end{verbatim}
\end{kframe}
\end{knitrout}

\section{Concluding remarks}\label{sec:Conc}
We are currently working on enhancements to \textbf{MESgenCov}.
Ultimately, we would like to make it easy to use
data sets from other application domains, and to make it
easier for a user to use other univariate  models than the one we provide.
Finally, we hope to eventually have a seamless integration
with algorithms for the MESP.

\section*{Acknowledgments} The authors are very grateful to Dr. Martin Shafer and Robert Larson for helping us gain access to the NADP/NTN data in a convenient form.

\section*{Financial and Ethical disclosures}
J. Lee was funded by the
Air Force Office of Scientific Research (Complex Networks program), FA9550-19-1-0175.
H. Al-Thani was funded by the Qatar National Research Fund (Graduate Sponsorship Research Award), GSRA4-2-0526-17114. The authors declare that they have no conflict of interest.

\bibliographystyle{alpha}
\bibliography{MESgenCov}

\newpage

\section{Appendices}

\subsection*{\bf Appendix A: NADP/NTN Data Descriptions}\label{appA}

\centering \includegraphics[width=\textwidth]{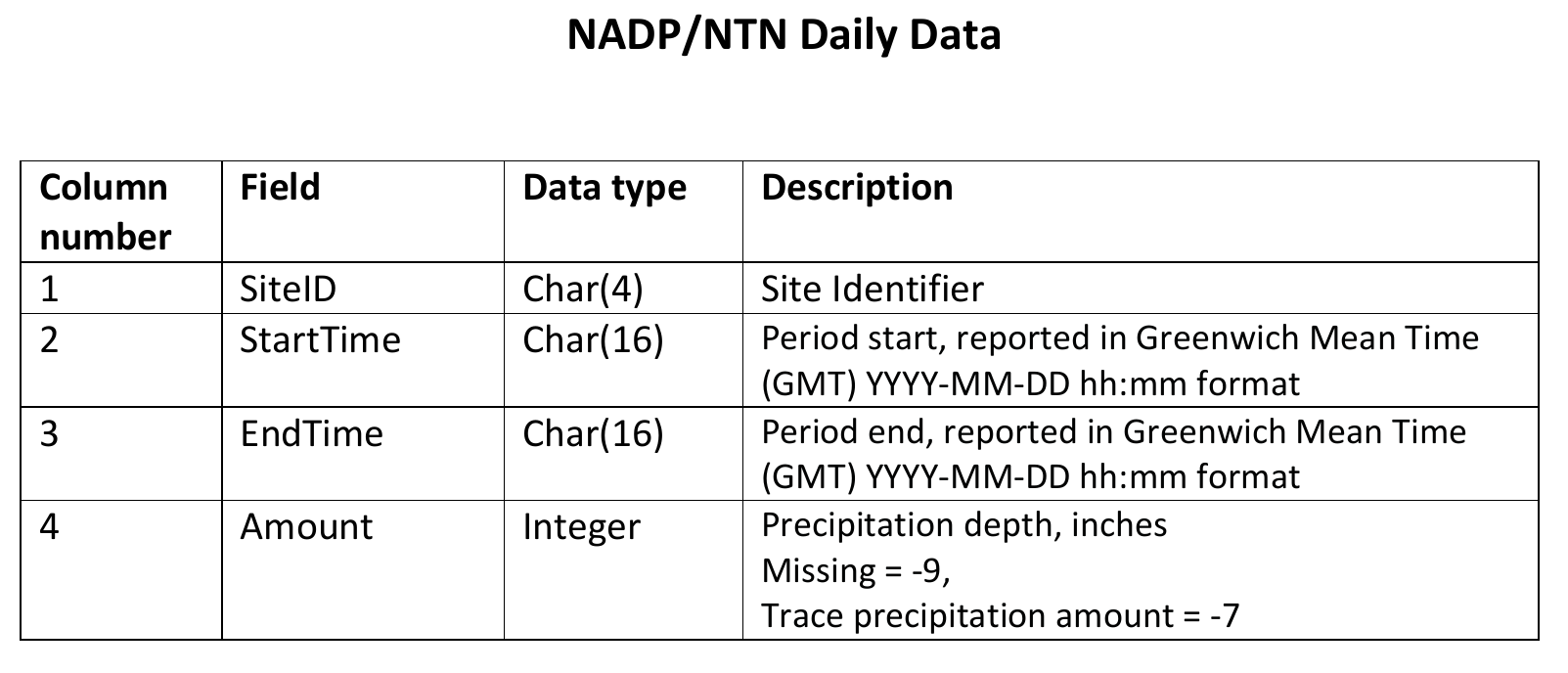}

\centering \includegraphics[width=\textwidth]{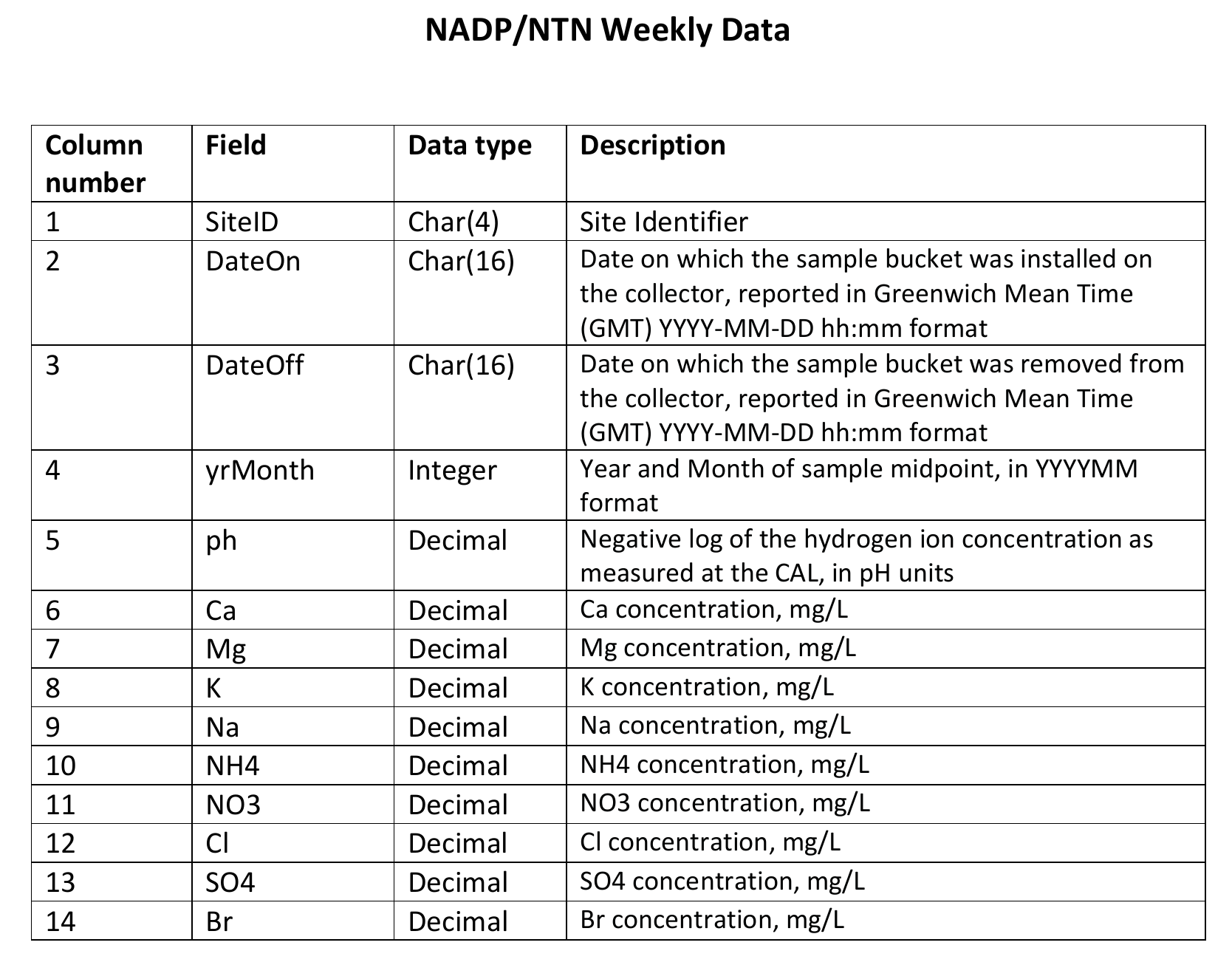}

\subsection*{\bf Appendix B: Geographic Split}\label{appB}
%\color{blue} Old image was based on an old four way split, this is what's currently being used by the code\color{black}
\begin{center}
\includegraphics[width = 1.0\textwidth]{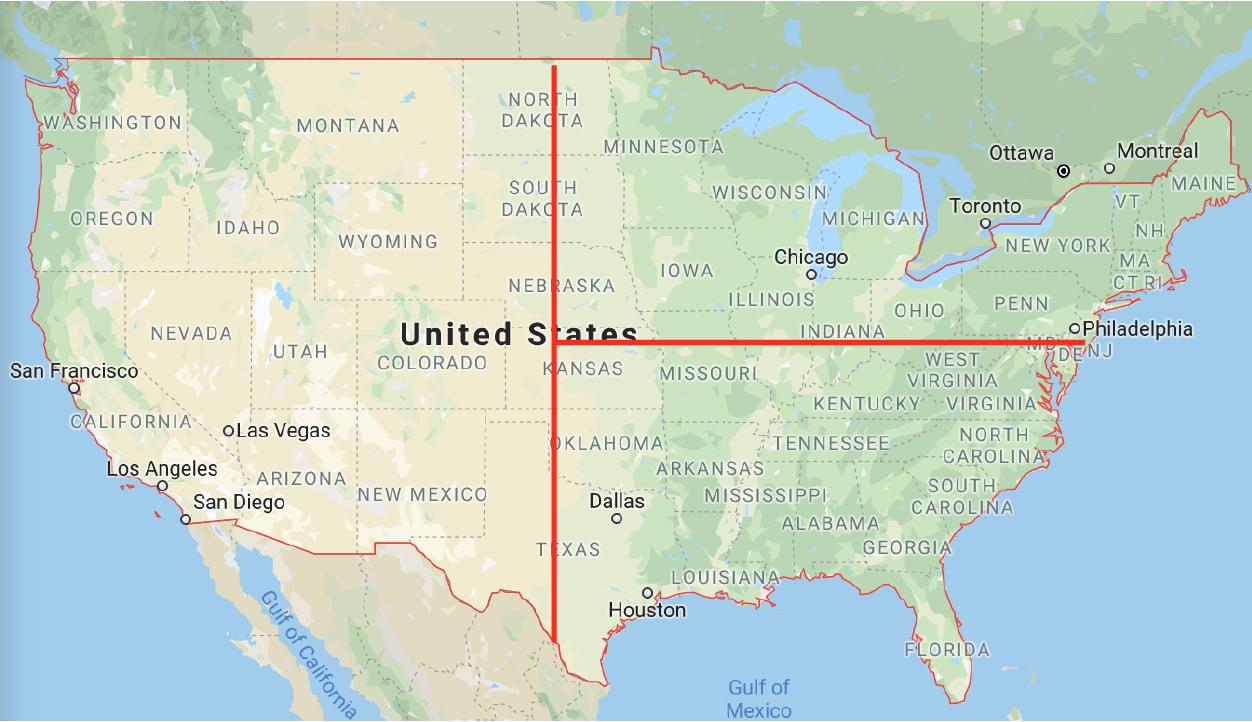}
\end{center}

\subsection*{\bf Appendix C: Internal Covariance Matrices Site Lists}\label{appC}
\begin{knitrout}
\definecolor{shadecolor}{rgb}{0.969, 0.969, 0.969}\color{fgcolor}\begin{kframe}
\begin{alltt}
\hlcom{#Sites with maximum distance data sets, get 50 sites}
\hlkwb{>}  \hlstd{maxd1}  \hlkwb{<-} \hlkwd{maxDistSites}\hlstd{(}\hlstr{"01/01/86 00:00"}\hlstd{,}\hlstr{"12/31/94 00:00"}\hlstd{,}\hlnum{50}\hlstd{,}
             \hlnum{200}\hlstd{,}\hlstr{"SO4"}\hlstd{,}\hlnum{1}\hlstd{)}
\hlkwb{>}  \hlstd{maxd2}  \hlkwb{<-} \hlkwd{maxDistSites}\hlstd{(}\hlstr{"01/01/07 00:00"}\hlstd{,}\hlstr{"12/31/14 00:00"}\hlstd{,}\hlnum{50}\hlstd{,}
             \hlnum{230}\hlstd{,}\hlstr{"SO4"}\hlstd{,}\hlnum{1}\hlstd{)}
\hlkwb{>}  \hlstd{maxd3}  \hlkwb{<-} \hlkwd{maxDistSites}\hlstd{(}\hlstr{"01/01/07 00:00"}\hlstd{,}\hlstr{"12/31/14 00:00"}\hlstd{,}\hlnum{50}\hlstd{,}
             \hlnum{230}\hlstd{,}\hlstr{"NO3"}\hlstd{,}\hlnum{1}\hlstd{)}
\hlkwb{>}  \hlstd{maxd4}  \hlkwb{<-} \hlkwd{maxDistSites}\hlstd{(}\hlstr{"01/01/07 00:00"}\hlstd{,}\hlstr{"12/31/14 00:00"}\hlstd{,}\hlnum{50}\hlstd{,}
             \hlnum{230}\hlstd{,}\hlstr{"Na"}\hlstd{,}\hlnum{1}\hlstd{)}
\hlkwb{>}  \hlstd{maxd5}  \hlkwb{<-} \hlkwd{maxDistSites}\hlstd{(}\hlstr{"01/01/07 00:00"}\hlstd{,}\hlstr{"12/31/14 00:00"}\hlstd{,}\hlnum{50}\hlstd{,}
             \hlnum{230}\hlstd{,}\hlstr{"ph"}\hlstd{,}\hlnum{1}\hlstd{)}
\end{alltt}
\end{kframe}
\end{knitrout}

%\includepdf[pages=-,pagecommand={},width=\textwidth]{ntn-daily-Meta.pdf}

%\subsection{Description of weekly data}

%\includepdf[pages=-,pagecommand={},width=\textwidth]{ntn-weekly-Meta.pdf}

\end{document}